\documentclass[a4paper,11pt]{article}
\usepackage{amsmath,amssymb}
\usepackage{graphics,graphicx}
\usepackage{booktabs}
\usepackage{hyperref}
\usepackage{a4wide}
\usepackage{cite}

\numberwithin{equation}{section}

\begin{document}
\thispagestyle{empty}
\begin{flushright}
\end{flushright}
\vspace{8mm}
\begin{center}
{\LARGE\bf On genus-$0$ invariants of Calabi-Yau hybrid models}
\end{center}
\vspace{10mm}
\begin{center}
  {\large David Erkinger\footnote{{\tt daviderkinger@aon.at}}, Johanna Knapp\footnote{{\tt johanna.knapp@unimelb.edu.au}}${}^{\dagger}$}
\end{center}
\vspace{3mm}
\begin{center}
  ${}^{\dagger}$ {\em School of Mathematics and Statistics, The University of Melbourne\\
  Parkville, VIC, 3010, Australia}
\end{center}
\vspace{15mm}
\begin{abstract}
  \noindent We compute genus zero correlators of hybrid phases of Calabi-Yau gauged linear sigma models (GLSMs), i.e.~of phases that are Landau-Ginzburg orbifolds fibered over some base. These correlators are generalisations of Gromov-Witten and FJRW invariants. Using previous results on the structure of the of the sphere- and hemisphere partition functions of GLSMs when evaluated in different phases, we extract the $I$-function and the $J$-function from a GLSM calculation. The $J$-function is the generating function of the correlators. We use the field theoretic description of hybrid models to identify the states that are inserted in these correlators. We compute the invariants for examples of one- and two-parameter hybrid models. Our results match with results from mirror symmetry and FJRW theory.  
\end{abstract}
\newpage
\setcounter{tocdepth}{1}
\tableofcontents
\setcounter{footnote}{0}
%%%%%%%%%%%%%%%%%%%%%%%%%%%%%%%%%%%%%%%%%%%%%%%%%%%%%%%%%%%%%%%%%%%%%%%
\section{Introduction}
The swampland program has led to renewed interest in classic topics of string theory such as Calabi-Yau compactifications and the associated moduli spaces. In this work we will focus on the stringy K\"ahler moduli space and on type II string theory on Calabi-Yau threefolds. The best studied regions in the K\"ahler moduli space are those near large volume points. While in the Calabi-Yau case every point in the moduli space corresponds to some worldsheet conformal field theory, moving away from the large volume regime is usually difficult as one typically lacks a concrete realisation of the worldsheet theory. Notable exceptions are Landau-Ginzburg points that have been studied in depth in the literature. A more general and less studied class of models are hybrid theories. They naturally appear at limiting points in the stringy K\"ahler moduli spaces associated to Calabi-Yaus. Hybrids can be formulated as fibrations of Landau-Ginzburg models over some base. Such theories have been discussed before, for instance in \cite{Bertolini:2013xga,Bertolini:2018now}, where in particular the state spaces were analysed. The aim of this work is to compute correlation functions in these theories that are captured by the topological sector. Concrete examples will focus on models with one and two parameters. 

The worldsheet theory we consider is the A-model of topological string theory, i.e.~a topological sigma model coupled to topological gravity. The correlation functions are characterised by inserting field operators and their gravitational descendants on the (closed) string worldsheet given by a Riemann surface of genus $g$. The correlation functions can be written in terms of integrals over the moduli space of the configuration of Chern classes of suitably chosen (orbi-)bundles over the insertion points. The techniques we are using will give us access to genus zero invariants. In the case of non-linear sigma models these correlators are the Gromov-Witten invariants, in the case of Landau-Ginzburg orbifolds they define FJRW-invariants \cite{Fan:2007ba}. 

We aim at computing these invariants for the case of hybrid CFTs that arise as phases of Calabi-Yau gauged linear sigma models (GLSMs). In the mathematics literature, the FJRW theory of one-parameter hybrids and its connection to Gromov-Witten theory and a generalised Landau-Ginzburg/Calabi-Yau correspondence has been developed in \cite{MR3590512,MR3888782,MR3842378,MR4283594,Zhao:2019dba}. To our knowledge, the invariants for these specific examples have been defined but have not been written down explicitly. Making use of the mathematical structure that is already in place and results from supersymmetric localisation in GLSMs, we use the following approach to extract the invariants. The invariants have a generating function given by the $J$-function \cite{MR1653024}. The $J$-function is related to the $I$-function by a coordinate transformation on the K\"ahler moduli space. Interpreted in the context of mirror symmetry, this transformation is the mirror map. %The $I$-function can be extracted from the sphere or hemisphere partition function of the gauged linear sigma models, and were computed for one- and two-parameter hybrid models in \cite{MR1653024}.

The $I$-function and, via the coordinate transformation, the $J$-function can be extracted from the sphere or hemisphere partition functions of the GLSM. In \cite{Knapp:2020oba,Erkinger:2020cjt} it has been conjectured that these partition functions have a universal structure when evaluated in any phase. Among other building blocks of the universal expressions are the $I$- and $J$-functions. This has been shown to work in examples of Calabi-Yau GLSMs with phases that are non-linear sigma models, Landau-Ginzburg orbifolds and hybrids where the $I$- and $J$-functions were known. Assuming that the structure holds more generally, one can compute (conjectural) expressions for the $I$- and $J$-functions in more general contexts. For the hybrids that we consider in this article, this was done in \cite{Erkinger:2020cjt} by evaluating the sphere partition function in hybrid phases of GLSMs. 

The $I$- and $J$-functions have an expansion in terms of elements of the ``narrow'' state space of the hybrid theory. This is a subring of the $(a,c)$-chiral ring of the worldsheet theory. The state spaces of $\mathcal{N}=(2,2)$ hybrid CFTs that are Landau-Ginzburg orbifolds fibered over a (Fano) base manifold have been analysed in \cite{Bertolini:2013xga}. Given these results, we identify the elements of the narrow state space with the elements of certain twisted sectors of the hybrid model. Generalising a statement of \cite{Knapp:2020oba} for Landau-Ginzburg orbifolds, we propose that the coordinate transformation relating the $I$- and $J$-functions can be extracted from components of the $I$-functions related to $(a,c)$-ring elements of left/right R-charges $(-1,1)$. Once we have the coordinate transformation, we obtain the $J$-function whose structure is constrained by selection rules related to the $U(1)_A$ and $U(1)_V$ symmetries. This makes it possible to read off the invariants.

We note that there are also higher genus invariants for the models we consider. However, at present, these are not accessible by our methods, as the $I$- and $J$-functions we use are only defined at genus zero. Also, the known results from supersymmetric localisation in GLSMs do not provide information beyond genus one. There has, however, been recent progress in the mathematics literature on the definition of GLSMs at higher genus and on the computation of higher genus invariants without mirror symmetry that may make it possible to tackle higher genus computations in the present framework, see eg.\cite{Fan:2015vca,Clader:2017cpe,Guo:2017scx,Guo:2018mol}. 

In this article, perform the calculation outlined above for one-parameter hybrid models that arise in the small volume regime of one-parameter complete intersection Calabi-Yaus in toric ambient spaces. We also apply these techniques to two two-parameter hybrid models and conjecture generalisations of hybrid FJRW invariants for them.

As non-trivial cross checks we show that our results are compatible with mirror symmetry by reinterpreting the coordinate transformation between the $I$- and the $J$-function as the mirror map and establishing a connection between components of the $J$-function and the mirror Yukawa couplings. Such an approach has been applied previously in \cite{Huang:2006hq}. For one of our examples we can also make a connection with instanton numbers previously computed in \cite{Sharpe:2012ji}.

The article is organised as follows. In Section~\ref{sec-jfun} we take the worldsheet perspective and discuss the definition of the $J$-function as the generating function of genus zero correlators. We recall some basic factorisation properties \cite{Witten:1989ig,Dijkgraaf:1990qw,Dijkgraaf:1990nc} and selection rules. Then we specialise to hybrid theories and state some further properties of the correlators that are specific for hybrid models\cite{MR3590512}. Furthermore we briefly remind the reader of the connection between the GLSM sphere partition function and the $I$-function for the case of hybrids\cite{Erkinger:2020cjt}. In Section~\ref{sec-hyb} we follow \cite{Bertolini:2013xga} to give a review of $\mathcal{N}=(2,2)$ hybrid CFTs and their state spaces. We then propose a physics characterisation of ``narrow'' states and show that this coincides with the mathematics definition for the cases where the latter is known. Furthermore we identify those elements of the state space that encode the transformation between the $I$-function and the $J$-function. Sections~\ref{sec-1par} and~\ref{sec-2par} are dedicated to computing hybrid FJRW invariants in one- and two-parameter examples, respectively. We end with discussing open questions and further research directions. \\\\
{\bf Acknowledgements:} We would like to thank Alessandro Chiodo, Ilarion Melnikov, Robert Pryor, Mauricio Romo, Emanuel Scheidegger, Thorsten Schimannek, and Eric Sharpe for helpful discussions and collaboration on related projects. JK thanks MATRIX Institute and Sorbonne Universit\'e for hospitality.
% remove if required by the referee
We also would like to thank the anonymous referee for helpful comments and for providing an explanation of the selection rule (\ref{rule2}) that has been added to a revised version of the article. DE was supported by the Austrian Science Fund (FWF): [P30904-N27]. JK is supported by the Australian Research Council Discovery Project DP210101502 and the Australian Research Council Future Fellowship FT210100514.
%%%%%%%%%%%%%%%%%%%%%%%%%%%%%%%%%%%%%%%%%%%%%%%%%%%%%%%%%%%%%%%%%%%%%%%%
\section{The $J$-function in $\mathcal{N}=(2,2)$ Calabi-Yau CFTs}
\label{sec-jfun}
In this section, we recall the definition of the $J$-function as the generating function for certain genus zero correlation functions of a topologically twisted $\mathcal{N}=(2,2)$ superconformal field theory coupled to topological gravity. This has been pioneered in \cite{Witten:1988xj,Witten:1989ig,Verlinde:1990ku,Dijkgraaf:1990nc}, see also \cite{Dijkgraaf:1990qw} for an excellent review. 
%%%%%
\subsection{Model-independent structure of the $J$-function}
The following discussion will use the notation of \cite{Knapp:2020oba} where more details can be found. First, we have to introduce some more background on the structure of the theory.

We consider an $\mathcal{N}=(2,2)$ superconformal field theory with central charge $c=3\hat{c}$. We will mostly focus on the case of Calabi-Yau threefolds, i.e.~$\hat{c}=3$. The states $\phi_i$ are elements of the (anti-)chiral rings of the conformal field theory. In our case, the relevant states are elements of the $(a,c)$-chiral ring: $\phi_i\in\mathcal{H}^{(a,c)}$ of dimension $\mathrm{dim}\mathcal{H}^{(a,c)}=2h+2$. Those $\phi_j$ ($j=1,\ldots,h$) with left- and right R-charges $(q,\overline{q})=(-1,1)$ parameterise marginal deformations of the $A$-twisted theory. We denote the associated deformation parameters by $t_j$. By $tt^*$-geometry \cite{Cecotti:1991me}, we can choose a set of flat coordinates on the moduli space of marginal deformations.

We define a set of basis elements $e_k$ ($k=0,\ldots,h$), where $e_0={\bf 1}$ is the unique state with $(q,\overline{q})=(0,0)$ and the $e_{k>0}$ are associated to the marginal deformations. The topological metric is given by $\eta_{ab}=\langle \phi_a,\phi_b\rangle$. This also defines a pairing on the state space. The dual basis elements $e^k$ ($k=0,\ldots,h$) are associated to the $h$ states of R-charges $(q,\overline{q})=(-2,2)$ for $k>0$ and the unique element with $(q,\overline{q})=(-3,3)$ for $k=0$, respectively.

%After coupling the theory to topological gravity, one obtains for each $\phi_i$ a tower of gravitational descendants. We denote by $\tau_n(\phi_a)$, the $n$-th gravitational descendant of $\phi_a$.

To understand the $J$-function, we need to couple the theory to two-dimensional topological gravity. In addition to the states $\phi_a$ there are further gravitational states $\tau_n(\phi_a)$ where $n\geq0$ and $\tau_0(\phi_a)=\phi_a$. The $\tau_n$, whose physics construction can be found in\footnote{The notation there was $\sigma_n$ rather than $\tau_n$.} \cite{Witten:1989ig,Dijkgraaf:1990nc}, can be understood as follows. Consider a string worldsheet with operator insertions, i.e.~a Riemann surface $C$ of genus $g$ with $m$ punctures, and the corresponding moduli space $\mathcal{M}_{g,m}$. Associate a line bundle, or more generally an orbibundle, $\mathcal{L}_i$ to the $i$-th marked point. The respective $\tau_n$ is then defined as\footnote{The the mathematics literature, these are also referred to as $\psi$-classes.} $\tau_n=c_1(\mathcal{L}_i)^n$. The most general correlators of the theory are
\begin{equation}
  \label{gencorr}
  \langle\tau_{n_1}(\phi_{j_1})\ldots\tau_{n_m}(\phi_{j_m})\rangle_{g,m}.
\end{equation}
%where $g$ is the genus of the worldsheet.
They can be defined as integrals over the moduli space. Schematically, this looks as follows:
\begin{equation}
  \label{intcorr}
  \langle\tau_{n_1}(\phi_{j_1})\ldots\tau_{n_m}(\phi_{j_m})\rangle_{g,m}=\int_{\overline{\mathcal{M}}_{g,m}}\xi\:\:\prod_{i=1}^mc_1(\mathcal{L}_i)^{n_i}\phi_{j_i},
\end{equation}
where $\overline{\mathcal{M}}_{g,m}$ is a suitable compactification of $\mathcal{M}_{g,m}$ and $\xi$ is the virtual fundamental class. A rigorous mathematical definition and construction of these objects is known in the case of Gromov-Witten theory (see eg.~\cite{Cox:2000vi,Hori:2003ic} for textbook accounts), FJRW theory \cite{Fan:2007ba,Fan:2007vi} and FJRW theory for certain examples of hybrids \cite{MR3590512}.  

The correlators are subject to the following $U(1)$ selection rule that comes from the $U(1)_A$-symmetry of the A-twisted theory:
\begin{equation}
  \label{rule1}
  \sum_{a=1}^m(n_a+\overline{q}_{j_a})=(\hat{c}-3)(1-g)+m.
\end{equation}

They furthermore satisfy various factorisation properties \cite{Witten:1989ig,Dijkgraaf:1990qw,Dijkgraaf:1990nc} which allows to reduce them further. These conditions have been formalised in mathematics for the case of Gromov-Witten invariants and FJRW invariants in the references given above.  Let ${\bf 1}=\mathrm{\tau}_0({\bf 1})$ be the identity operator corresponding to an insertion of the state with $(q,\overline{q})=(0,0)$. Insertions of this operator can be removed by the {\em string equation} (or puncture equation):
\begin{equation}
  \label{stringeq}
  \langle \tau_{n_1}(\phi_{j_1})\ldots\tau_{n_m}(\phi_{j_m}){\bf 1}\rangle_{g,m+1}=\sum_{i=1}^m\langle \tau_{n_i-1}(\phi_{j_i})\tau_{n_1}(\phi_{j_1})\ldots\widehat{\tau_{n_i}(\phi_{j_i})} \ldots \tau_{n_m}(\phi_{j_m}) \rangle_{g,m},
\end{equation}
where the hat means omitting the corresponding insertion. Furthermore, an insertion of $\tau_1({\bf 1})$, corresponding to inserting a dilaton operator, is removed using the {\em dilaton equation}
\begin{equation}
  \label{dilatoneq}
  \langle \tau_{n_1}(\phi_{j_1})\ldots\tau_{n_m}(\phi_{j_m})\tau_1({\bf 1})\rangle_{g,m+1}=(2g-2+m)\langle \tau_{n_1}(\phi_{j_1})\ldots\tau_{n_m}(\phi_{j_m})\rangle_{g,m}.
  \end{equation}
Correlation functions also satisfy topological recursion relations. As we will not need them, we refrain from stating them.

Now we have all the ingredients to state the $J$-function for an $\mathcal{N}=(2,2)$ Calabi-Yau SCFT. It can be written as
\begin{equation}
J=e_0+\sum_{i=1}^ht_ie_i+\sum_{\overline{q}_c<\hat{c}-1}\sum_{n=1}^{\infty}\sum_{k_1,\ldots,k_h\geq 0}\prod_{i=1}^h\frac{(t_i)^{k_i}}{k_i!}\langle\tau_n(\phi_c)\prod_{j=1}^h\phi_j^{k_j}\rangle_{0,|k|+1}\eta^{cb}e_b. \label{cft-jfunction}
\end{equation}
%%%%%%%%
\subsection{The $J$-function for hybrid models}
\label{sec-jhyb}
The conditions stated thus far can be understood purely in terms of the two-dimensional field theory coupled to topological gravity. There are further conditions that are more specific to the concrete realisation of such a theory and consequently the properties of the insertions $\phi_{j_i}$. In geometric settings described by non-linear sigma models with Calabi-Yau target $X$, the $(c,c)$- and $(a,c)$-states can be described in terms of the cohomology of $X$. In the case of Landau-Ginzburg orbifolds, the $(c,c)$- and $(a,c)$-elements correspond to states in the untwisted and twisted sectors of the theory. 

As we will discuss in more detail in Section~\ref{sec-hyb}, we consider hybrid theories with are Landau-Ginzburg orbifolds fibred over a base $B$. Schematically, a state $\phi$ has the following structure:
\begin{equation}
  \label{statestructure}
  \phi=\phi_B\otimes\phi_{\delta},
\end{equation}
where $\phi_{B}$ is represented by a cohomology element of the base $B$ and $\phi_{\delta}$ is an element of the $\delta$-th twisted sector in the fibre Landau-Ginzburg orbifold.

An important special case, related to the narrow sectors to be proposed in Section~\ref{sec-hyb}, are states for which $\phi_{\delta}={\bf 1}_{\delta}$. i.e.~states which have the identity/vacuum in the fibre direction. Such states can be characterised by the cohomology of the base $B$. As special cases, there can be states $\tilde{\phi}$ that can be expanded as $\tilde{\phi}=\sum_i\alpha_i e_{H_i}$ with $e_{H_i}\in H^2(B)$.  We can take an integral curve class $\beta=\sum_i\beta_iH_i$ with $H_i\in H_2(B)$ so that $\int_{\beta}\tilde{\phi}=\sum\alpha_i\beta_i$. The correlators (\ref{gencorr}) get an additional label\footnote{We abuse notation here. We hope that the distinction of $\beta_i$ and the two-cycle $\beta$ is clear from the context.} $\beta=\{\beta_i\}$ that keeps track of the choice of $\beta$. Insertions of this type of operators can be removed by the {\em divisor equation} (or divisor axiom) \cite{MR1291244} (see also \cite{Cox:2000vi} or \cite{Hori:2003ic} for reviews)
\begin{equation}
  \label{divisoreq}
  \langle \tau_{n_1}(\phi_{j_1})\ldots\tau_{n_m}(\phi_{j_m})\tilde{\phi}\rangle_{g,m+1,\beta}=\left(\int_{\beta}\tilde{\phi}\right)\langle \tau_{n_1}(\phi_{j_1})\ldots\tau_{n_m}(\phi_{j_m})\rangle_{g,m,\beta}+\ldots,
\end{equation}
where '$\ldots$' denotes further terms that are present whenever $\tau_n(\phi_j)$ with $n>0$ are inserted. Since the only insertions that can occur in our setting will turn out to be $\tau_{0}(\phi_{j_i})$ and $\tau_1({\bf 1})$, the latter of which can be removed by the dilaton equation, these extra terms never show up for the invariants we compute in this work.

Note that in the case of Gromov-Witten theory, the calculation of the genus zero-invariants reduces to computing invariants where all the possible insertions are of the type $\tilde{\phi}$. For $\beta>0$, the divisor equation makes it possible to remove all insertions so that correlators without any gravitational states are of the form $\langle\cdot\rangle_{0,0,\beta}$. In the case of FJRW theory there is no divisor axiom. 

In the definition of the $J$-function, an additional summation over all $\beta$ has to be added. The $J$-function for hybrids therefore has the following expansion \cite{MR3590512}:
\begin{equation}
  J=e_0+\sum_{i=1}^ht_ie_i+\sum_{\beta}\sum_{\overline{q}_c<\hat{c}-1}\sum_{n=1}^{\infty}\sum_{k_1,\ldots,k_h\geq 0}\prod_{i=1}^h\frac{(t_i)^{k_i}}{k_i!}\langle\tau_n(\phi_c)\prod_{j=1}^h\phi_j^{k_j}\rangle_{0,|k|+1,\beta}\eta^{cb}e_b. \label{hyb-jfunction}
\end{equation}

In the context of Landau-Ginzburg orbifolds and hybrids, there is a further selection rule associated to the action of the orbifold group. To understand its origin and possible generalisations, we have to make a few technical remarks on the nature of hybrid models that will be elaborated on further in the subsequent sections. In the physics literature on hybrid models, the base $B$ is often assumed to be smooth Fano variety. However, it can happen that the orbifold group of the Landau-Ginzburg fibre also acts on $B$, albeit in a trivial fashion. In this context, ``trivial'' means that all the scalar fields in the hybrid theory, which parameterise the base $B$, may transform with charge $N$, or multiples thereof, under the action of an orbifold group $\mathbb{Z}_N$ associated to the Landau-Ginzburg fibre. This is different from the group not acting at all on $B$ in which case the associated scalar fields would have charge $0$. The correct mathematical framework to understand this is in terms of gerbes. See for instance \cite{Sharpe:1999pv,Sharpe:1999xw,Pantev:2005zs,Pantev:2005wj} for references that are useful in the present context. In the field theoretic characterisation of the states \cite{Bertolini:2013xga} to be reviewed in Section~\ref{sec-hyb} and in the mathematics formulation of the state space \cite{MR3590512} (see in particular remarks 4.1.4 and 4.1.5 therein) one can formally replace the gerbe by a smooth variety. However, the fact that there is some action of a group has a non-trivial effect, for instance for certain selection rules associated to the action of the orbifold group.

In \cite{MR3590512}, a corresponding selection rule has been stated for a class of such ``gerby'' one-parameter hybrids. In more general settings, we expect a selection rule of the following form for the hybrid correlator $\langle\tau_{n_1}(\phi_{j_1})\ldots\tau_{n_m}(\phi_{j_m})\rangle_{g,m,\beta}$
\begin{equation}
  \label{rule2}
  2g-2+m-q_B(\beta)-\sum_{i=1}^mq_{\delta_{j_i}}=0\mod N,
\end{equation}
where $q_B(\beta)$ accounts for any action of the orbifold group on the base. The term $q_{\delta_{j_i}}$ accounts for the transformation property of the fibre state $\phi_{\delta}$ in (\ref{statestructure}). For the special case where $\phi_{\delta}={\bf 1}_{\delta}$ we find $q_{\delta_{j_i}}=\delta_{j_i}$. From the physics perspective, this selection rule can be traced back to the fact that the orbifold group acts on the Hilbert space of the quantum theory. It should thus be interpreted as originating from a generalisation of the quantum symmetry for Landau-Ginzburg orbifolds \cite{Vafa:1989ih} to the case of hybrids. In Sections~\ref{sec-1par} and~\ref{sec-2par}, we will state the explicit form of this selection rule for all our examples and we will show that it is consistent with the structure of the $J$-functions of these models. It would be interesting to explore this further and to give a physics derivation for the general form of this rule, based on the defining data of the states.
% remove below here if the referee does not agree

From the perspective of FJRW theory for hybrids, the selection rule has the following origin\footnote{We are grateful to the anonymous referee for supplying an explanation.}. Given a punctured Riemann surface $C$, the construction of the moduli space $\overline{\mathcal{M}}_{g,m}$ requires the notion of orbifold stable curves $\mathcal{C}$ where the marked points and nodes can be orbifold points. One can equip $\mathcal{C}$ with a degree $2g-2$ sheaf $\omega_{\mathrm{log}}$ of logarithmic differential forms with simple poles only at nodes and marked points. Then one constructs an orbifolded line bundle $L$ over $\mathcal{C}$ that satisfies
\begin{equation}
  L^{\otimes N}\cong\omega_{\mathrm{log}}\otimes \mathcal{O}\left(-\sum_{i=1}^mq_i[x]_i\right)\otimes\mathcal{O}(-q_B(\beta)).
\end{equation}
The second term on the righthand side accounts for the information of the twisted sector states inserted at the punctures, the third term comes from the pullback of the line bundle $\mathcal{O}(-1)$ on $B$ to the punctured Riemann surface\footnote{$L$ and $\mathcal{C}$ are part of the data required to define the virtual fundamental cycle and the virtual fundamental class that are needed to make the definition (\ref{intcorr}) of the correlators precise. We refer to the mathematics literature for the complete picture.}. The selection rule follows from this property. For the one-parameter hybrids discussed in Sections~\ref{sec-k1} and \ref{sec-m1} this coincides with \cite{MR3590512}. For genus zero, similar results for Landau-Ginzburg models have been obtained by analysing the instanton moduli space of the GLSM \cite{Witten:1993yc}. 
%%%%%
\subsection{$J$-function via the GLSM}
We will compute FJRW invariants for certain hybrid models by computing the $J$-function via the $I$-function. As we have discussed, the $J$-function (\ref{cft-jfunction}) is a function $J(t)$ of coordinates $t$ which can be identified with the deformation parameters associated to the marginal deformations of the worldsheet CFT. The $I$-function $I(u)$ is related to the $J$-function via a change of frame and coordinates:
\begin{equation}
  \label{ijrel}
  J(t)=\frac{I(u(t))}{I_0(u(t))}.
\end{equation}
To understand this formula, we recall that, just like the $J$-function, the $I$-function can be expanded in terms of a basis of the $(a,c)$-ring. The component $I_0$ is the coefficient of $e_0$ while the coordinate transformation is encoded in the components associated to the marginal deformations with $(q,\overline{q})=(-1,1)$. For the Landau-Ginzburg orbifold case, this was discussed in \cite{Knapp:2020oba}. Below, we will find a straightforward generalisation for hybrid models.

In \cite{Jockers:2012dk}, it was first observed that the sphere partition function of a Calabi-Yau GLSM computes the K\"ahler potential on the associated K\"ahler moduli space. In \cite{Erkinger:2020cjt} it was shown that for a GLSM that has a phase that is a ``good'' hybrid with orbifold group $G$, the sphere partition function reduces to 
\begin{equation}
  \label{hybifun}
Z^{\textrm{hyb}}_{S^2}=C\sum_{\delta\in G}\int_B(-1)^{\mathrm{Gr}}\frac{\widehat{\Gamma}_{\delta}(H)}{\widehat{\Gamma}^*_{\delta}(H)}I_{\delta}(u(\mathsf{t}),H)\overline{I}_{\delta}(\overline{u}(\overline{\mathsf{t}}),H).
  \end{equation}
Here, $C$ is an undetermined normalisation factor, the sum $\delta\in G$ runs over the narrow sectors, $\mathrm{Gr}$ is a grading on the state space, the details of which we will not need for our discussion. Furthermore, $H\in H^2(B)$, $\widehat{\Gamma}$ denotes the Gamma class, and $\mathtt{t}$ is the FI-theta parameter of the GLSM. Finally, $I_{\delta}(u(\mathtt{t}),H)$ is the component of the $I$-function associated to the $\delta$-twisted sector that can be expanded further in powers of $H\in H^2(B)$. This structure has been used in \cite{Erkinger:2020cjt} to identify the $I$-functions of Calabi-Yau hybrid models.
%%%%%%%%%%%%%%%%%%%%%%%%%%%%%%%%%%%%%%%%%%%%%%%%
\section{Hybrid models and state spaces}
\label{sec-hyb}
In order to understand the components of the $I$-function and the $J$-function and the insertions of the correlators, we have to know more about the state space and in particular the $(a,c)$-rings of the hybrid models we are interested in. A certain class of ``good'' hybrid models that flow to an $\mathcal{N}=(2,2)$ CFT have been analysed in detail in \cite{Bertolini:2013xga}. The state space of these models has been computed and interpreted in the context of heterotic string compactifications. Compared to the Landau-Ginzburg orbifold case, computing these states is a rather non-trivial task and requires making use of spectral sequences whose differentials, roughly speaking, arise from the components of the supercharges in the base and fibre directions of the hybrid model.

For our purposes, we only need to identify the states corresponding to the elements of the $(a,c)$-ring that belong to the so-called narrow sector. Results from FJRW theory \cite{MR3590512,MR3888782} imply that these states should be in correspondence with elements of the cohomology of the base of the fibration. In this section, we propose a way to characterise narrow states in hybrid models and show that this narrow state space is indeed consistent with the mathematical formulation. The states in the narrow sectors have a much simpler structure than the generic states. The use of spectral sequences to compute the state space can be avoided, as the cohomology coming from the Landau-Ginzburg orbifold fibres is trivial in this case. To understand these states, we first have to recall the results of \cite{Bertolini:2013xga} before we can show how the state space simplifies in the proposed narrow sectors.  
%%%%%
\subsection{$\mathcal{N}=(2,2)$ hybrid theories}
We start by recalling the definition of a ``good'' hybrid model. We take a K\"ahler manifold $Y_0$ together with a superpotential $W$ such that
\begin{equation}
  \label{potentialcond}
  dW^{-1}(0)=B\subset Y_0,
  \end{equation}
where $B$ is some compact subset of complex dimension $d$. Condition (\ref{potentialcond}) is called the potential condition. Locally, the geometry can be modelled by the total space of a rank $n$ holomorphic vector bundle $X\rightarrow B$ which is denoted by $Y$. The low energy physics of the hybrid theory is encoded in the geometry of $Y$.

The field content of the hybrid model with $\mathcal{N}=(2,2)$ supersymmetry can be decomposed in terms of $\mathcal{N}=(0,2)$ chiral and anti-chiral multiplets. We denote\footnote{The translation to the familiar $\mathcal{N}=(2,2)$ notation is $(\theta^+,\overline{\theta}^+)\leftrightarrow (\theta,\overline{\theta})$, $(\theta^-,\overline{\theta}^-)\leftrightarrow(\theta',\overline{\theta}')$.} by $(\theta,\overline{\theta})$ the right-moving Grassmann coordinates and by $(\theta',\overline{\theta}')$ their left-moving counterparts. The left and right R-charges of $(\theta',\theta)$ are $(q,\overline{q})=(1,1)$, respectively. We define the covariant derivatives
\begin{equation}
  \mathcal{D}=\frac{\partial}{\partial\theta}+\overline{\theta}\overline{\partial}_{\overline{z}}, \qquad \overline{\mathcal{D}}=\frac{\partial}{\partial\overline{\theta}}+\theta\overline{\partial}_{\overline{z}},
\end{equation}
and analogously $(\mathcal{D}',\overline{\mathcal{D}}')$. There are bosonic $(0,2)$ chiral multiplets $Y^{\alpha}$ and chiral Fermi multiplets $\mathcal{X}^{\alpha}$, together with their anti-chiral counterparts, as follows:
\begin{align}
  Y^{\alpha}&=y^{\alpha}+\sqrt{2}\theta\eta^{\alpha}+\theta\overline{\theta}\overline{\partial}_{\overline{z}}y^{\alpha},& \overline{Y}^{\overline{\alpha}}&=\overline{y}^{\overline{\alpha}}-\sqrt{2}\overline{\theta}\overline{\eta}^{\overline{\alpha}}-\theta\overline{\theta}\overline{\partial}_{\overline{z}}\overline{y}^{\overline{\alpha}},\nonumber\\
  \mathcal{X}^{\alpha}&=\chi^{\alpha}+\sqrt{2}\theta H^{\alpha}+\theta\overline{\theta}\overline{\partial}_{\overline{z}}\chi^{\alpha},& \overline{\mathcal{X}}^{\overline{\alpha}}&=\overline{\chi}^{\overline{\alpha}}+\sqrt{2}\overline{\theta}\overline{H}^{\overline{\alpha}}-\theta\overline{\theta}\overline{\partial}_{\overline{z}}\overline{\chi}^{\overline{\alpha}},
  \end{align}
where $\alpha,\overline{\alpha}=1,\ldots,\mathrm{dim}Y$. The $H^{\alpha}$ and their conjugates are auxiliary fields which can be eliminated by the equations of motion.  Chiral and anti-chiral $(2,2)$ superfields can then be written as
\begin{equation}
  \mathcal{Y}^{\alpha}=Y^{\alpha}+\sqrt{2}\theta'\mathcal{X}^{\alpha}+\theta'\overline{\theta}'\partial_zY^{\alpha},\qquad\qquad \overline{\mathcal{Y}}^{\overline{\alpha}}=\overline{Y}^{\overline{\alpha}}-\sqrt{2}\theta'\overline{\mathcal{X}}^{\overline{\alpha}}-\theta'\overline{\theta}'\partial_z\overline{Y}^{\overline{\alpha}}.
\end{equation}
Due to the fibration structure, it is convenient to split the chiral scalars $y^{\alpha}$ as $y^{\alpha}=(y^I,\phi^i)$, where the $y^I$, $I=1,\ldots,d$, are the base coordinates and the $\phi^i$ with $i=1,\ldots,n$ are the fibre coordinates. Using these superfields, one can construct an action for the hybrid model. Since we do not need the explicit form, we refer to \cite{Bertolini:2013xga} for details. All we need are the transformation properties of the fields under supersymmetry. We denote the right-moving supersymmetry operators by $Q$ and $\overline{Q}$, respectively. The operator $\overline{Q}$ acts as follows:
\begin{align}
  [\overline{Q},y^{\alpha}]&=0, & [\overline{Q},\chi^{\alpha}]&=0, & [\overline{Q},\eta^{\alpha}]&=\overline{\partial}_{\overline{z}}y^{\alpha}, & [\overline{Q},H^{\alpha}]&=\overline{\partial}_{\overline{z}}\chi^{\alpha},\nonumber\\
  [\overline{Q},\overline{y}^{\overline{\alpha}}]&=-\overline{\eta}^{\overline{\alpha}}, & [\overline{Q},\overline{\chi}^{\overline{\alpha}}]&=\overline{H}^{\overline{\alpha}}, & [\overline{Q},\overline{\eta}^{\overline{\alpha}}]&=0, & [\overline{Q},\overline{H}^{\overline{\alpha}}]&=0. 
\end{align}
Here $[\cdot,\cdot]$ is a graded commutator. The action of $Q$ can be obtained by replacing chiral and anti-chiral degrees of freedom. The respective action of the left-moving supercharge $Q'$ can be obtained by switching left- and right-moving fermions $\chi^{\alpha}$ and $\eta^{\alpha}$, and similarly for $\overline{Q}'$. 

Integrating out the non-propagating degrees of freedom results in
\begin{equation}
  g_{\alpha\overline{\beta}}\overline{H}^{\overline{\beta}}=g_{\alpha\overline{\beta},\overline{\gamma}}\overline{\eta}^{\overline{\gamma}}\overline{\chi}^{\overline{\beta}}+W_{\alpha},
\end{equation}
where $g_{\alpha\overline{\beta}}(Y^{\alpha},\overline{Y}^{\overline{\alpha}})$ is the K\"ahler metric on $Y$, which can be used to raise and lower indices. Furthermore, $W_{\alpha}=\partial W/\partial Y^{\alpha}$, etc. It is also convenient to introduce the left-moving bosons
\begin{equation}
  \rho_{\alpha}=g_{\alpha\overline{\alpha}}\partial\overline{y}^{\overline{\alpha}}+\Gamma_{\alpha\gamma}^{\delta}\overline{\chi}_{\delta}\chi^{\gamma},
  \end{equation}
with the Christoffel symbol $\Gamma_{\alpha\gamma}^{\delta}$. One finds that $\overline{Q}$ decomposes as $\overline{Q}=\overline{Q}_0+\overline{Q}_W$, where $\overline{Q}_0$ and $\overline{Q}_W$ anticommute. The non-zero on-shell actions on the fields are
\begin{equation}
  [\overline{Q}_0,\overline{y}^{\overline{\alpha}}]=-\overline{\eta}^{\overline{\alpha}}, \quad [\overline{Q}_0,\eta^{\alpha}]=\overline{\partial}_{\overline{z}}y^{\alpha}, \quad [\overline{Q}_W,\overline{\chi}_{\alpha}]=W_{\alpha}, \quad [\overline{Q}_W,\rho_{\alpha}]=\chi^{\beta}W_{\beta\alpha}.
  \end{equation}
As long as $W=0$ there is an exact $U(1)_L\times U(1)_R$ R-symmetry that acts trivially on the bosons and in the following way on the fermions:
\begin{equation}
  \delta_L^0\eta=0,\quad \delta_L^0\chi=-i\epsilon\chi;\qquad\qquad \delta_R^0\eta=-i\epsilon\eta,\quad \quad \delta_R^0\chi=0.
\end{equation}
To promote this symmetry to a symmetry of the full hybrid model with $W\neq 0$, one requires the existence of a holomorphic Killing vector field $V$ on the geometry $(Y,g)$ such that $\mathcal{L}_VW=W$ where $\mathcal{L}$ is the Lie derivative. The associated symmetry acts non-chirally on the superfields:
\begin{equation}
  \delta_VY^{\alpha}=i\epsilon V^{\alpha}(Y),\quad \delta_V\overline{Y}^{\overline{\alpha}}=-i\epsilon\overline{V}^{\overline{\alpha}}(\overline{Y}),\qquad \delta_V\mathcal{X}^{\alpha}=i\epsilon V^{\alpha}_{,\beta}\mathcal{X}^{\beta},\quad \delta_V\overline{\mathcal{X}}^{\overline{\alpha}}=-i\epsilon\overline{V}^{\overline{\alpha}}_{,\overline{\beta}}\overline{\mathcal{X}}^{\overline{\beta}}.
\end{equation}
Given these restrictions, $\delta_{L,R}=\delta_{L,R}^0+\delta_V$ are classical symmetries of the action. For $U(1)_L$ to be anomaly-free, we further impose that $Y$ has trivial canonical bundle which implies that $B$ is Fano. The conserved $(U(1)_{L},U(1)_R)$-charges will be denoted by $(q,\overline{q})$, respectively.

To get well-defined R-symmetries in the IR SCFT one has to impose the additional condition that $V$ is a vertical vector field, i.e.~that $\mathcal{L}_V\pi^{\ast}(\omega)=0$ for all forms $\omega\in\Omega^{\bullet}(B)$ on $B$. Models that satisfy this condition are referred to as {\em good hybrids} and one can show that
\begin{equation}
  \label{killing}
  V=\sum_{i=1}^{n}q_i\phi^i\frac{\partial}{\partial \phi^i}+\mathrm{c.c.}, 
\end{equation}
where $0<q_i\leq\frac{1}{2}$ are rational. With these structures in place, the hybrid model is a Landau-Ginzburg model in the fibre fields $\phi^i$ on every local patch on $B$. Models that are not good hybrids have been named pseudo-hybrids \cite{Aspinwall:2009qy} and are related to singular CFTs. 

Our focus is on hybrid models arising in the context of type II string compactifications on Calabi-Yau threefolds. This means that the hybrid theory should flow in the IR to an SCFT with central charge $(c,\overline{c})=(9,9)$. Such theories require that all states in the (NS,NS) sector have integral R-charges $(q,\overline{q})$. To achieve this, one has to gauge the discrete symmetry induced by $U(1)_L$. If the fibre fields have fractional charges $q_i=\frac{n_i}{d_i}$ for some $n_i$ and $d_i$, this gives a $\mathbb{Z}_N$-orbifold with $N=\mathrm{lcm}(d_1,\ldots,d_n)$. 

As we have mentioned in Section~\ref{sec-jhyb}, we can have a slightly more general setup for a consistent hybrid model if we consider such an orbifold. We can now allow for the orbifold group to act on $B$, but this action has to be trivial, implying that the hybrid model should be formulated in terms of gerbes. However, even in this case, the R-charges of the base fields will be zero, which means that the formalism developed in \cite{Bertolini:2013xga} still applies without any modification. Therefore we can, for the sake of constructing the states in CFT, replace the ``gerby'' $B$ by an associated Fano manifold, consistent with what has also been observed in the mathematics literature. This is an example of a setup where the orbifold group of a hybrid model does not necessarily embed into the left R-symmetry group. In general, one expects that there can be hybrid models with more general discrete groups acting non-trivially on the fibre. In this case, there can be additional effects, as implied for instance by the selection rule (\ref{rule2}). We will also refer to such slightly more general models as good hybrids. We will elaborate some more on these subtle issues when we discuss concrete examples. 

%%%%%
\subsection{State space}
Once we have set up a good hybrid model, we have to determine the massless spectrum. This accounts for the degrees of freedom of the SCFT in the IR. What we are primarily interested in are massless states in the (R,R)- and (NS,NS)-sectors of the compactified theory because they correspond to the ground states of the internal CFT, i.e.~the elements of the chiral rings. Due to $\{Q,\overline{Q}\}=2\overline{L}_0$, the states with zero right-moving energy are in the cohomology of $\overline{Q}$. Hence, computing these cohomologies is the first step towards identifying the states we want. Those states that are also in the cohomology of $\overline{Q}'$ and $Q'$ correspond, via spectral flow, to the elements of the $(c,c)$- and $(a,c)$-rings, respectively. 

If we consider a full-fledged string compactification we must further impose left- and right-moving GSO projections. In the left-moving sector, the $\mathbb{Z}_2$ determining the GSO projection and the $\mathbb{Z}_N$ of the orbifold group can be combined into a $\mathbb{Z}_{2N}$. States in the $k$-th twisted sector satisfy $\phi(e^{2\pi i}z,e^{-2\pi i}\overline{z})=(e^{\pi i J_0})^k\phi(z,\overline{z})$ ($k=0,\ldots,2N-1$) where $J_0$ is the conserved $U(1)_L$-charge. In this framework, states in the (R,R)-sector arise for even $k$.

All the generators of the left-moving supersymmetry algebra are $\overline{Q}$-closed and thus give a well-defined action on the cohomology $H_{\overline{Q}}$. The $U(1)_L$-current $J_L$, the left-moving energy-momentum tensor $T$, and the left-moving supercurrents are
\begin{align}
  \label{currents}
  J_L&=\chi^{\beta}(V^{\alpha}_{\phantom{\alpha},\beta}-\delta^{\alpha}_{\beta})\overline{\chi}_{\alpha},\nonumber\\
  T&=-\partial y^{\alpha}\rho_{\alpha}-\frac{1}{2}\left(\overline{\chi}_{\alpha}\partial_z\chi^{\alpha}+\chi^{\alpha}\partial_z\overline{\chi}_{\alpha}\right)-\frac{1}{2}\partial_z\left(\chi^{\beta}\overline{\chi}_{\alpha}V^{\alpha}_{\phantom{\alpha},\beta}-v^{\alpha}\rho_{\alpha} \right),\nonumber\\
  G^{+}&=i\sqrt{2}\left(\overline{\chi}_{\alpha}\partial_zy^{\alpha}-\partial_z(\overline{\chi}_{\alpha}V^{\alpha})\right),\nonumber\\
  G^-&=i\sqrt{2}\chi^{\alpha}\rho_{\alpha}.
  \end{align}
The left-moving fields satisfy the following free-field OPEs:
\begin{equation}
  \label{freeope}
  y^{\alpha}(z)\rho_{\beta}(w)\sim\frac{1}{z-w}\delta^{\alpha}_{\beta}, \qquad \chi^{\alpha}(z)\overline{\chi}_{\beta}(w)\sim\frac{1}{z-w}\delta^{\alpha}_{\beta}.
\end{equation}
Using these OPEs and the explicit form (\ref{killing}) of $V$, one finds the left central charge
\begin{equation}
  c=3d+3\sum_{i=1}^n(1-2q_i),
  \end{equation}
which is the expected result for a Landau-Ginzburg model fibered over a base manifold. Given this information, the left and right R-charges and the conformal weights of the base and fibre fields are
\begin{equation}
  \begin{array}{c|cccc|cccc}
    &y^I&\rho_I&\chi^I&\overline{\chi}_I&\phi^i&\rho_i&\chi^i&\overline{\chi}_i\\
\hline
    q&0&0&-1&1&q_i&-q_i&q_i-1&1-q_i\\
    \overline{q}&0&0&0&0&q_i&-q_i&q_i&-q_i\\
    2h&0&2&1&1&q_i&2-q_i&1+q_i&1-q_i
    \end{array}
\end{equation}

The elements of the $\overline{Q}$-cohomology are graded with respect to their energy $E$ and their left R-charge $q$. Since we are working in an orbifolded theory, we have to determine $(q,\overline{q},E)$ for each twisted sector $k$. In particular, we have to determine these charges for the ground states $|k\rangle$ in each sector. If we consider the Ramond ground states in the right-moving sector, only the zero-modes contribute to the $\overline{Q}$-cohomology. In the left-moving sector, oscillator modes can contribute. In the $k$-th twisted sector the oscillator expansions of the left-moving fields are
\begin{align}
  \label{modes}
  y^{\alpha}(z)&=\sum_{r\in\mathbb{Z}}y_r^{\alpha}z^{-r-h_{\alpha}}& \chi^{\alpha}(z)&=\sum_{r\in\mathbb{Z}-\tilde{\nu}_{\alpha}}\chi_r^{\alpha}z^{-r-\tilde{h}_{\alpha}}\nonumber\\
  \rho_{\alpha}(z)&=\sum_{r\in\mathbb{Z}+\nu_{\alpha}}\rho_{\alpha r}z^{-r+h_{\alpha}-1}& \overline{\chi}_{\alpha}(z)&=\sum_{r\in\mathbb{Z}+\tilde{\nu}_{\alpha}}\overline{\chi}_{\alpha r}z^{-r+\tilde{h}_{\alpha}-1},
\end{align}
with 
\begin{equation}
  \label{nudef}
  \nu_{\alpha}=\frac{kq_{\alpha}}{2}\mod 1, \quad \tilde{\nu}_{\alpha}=\frac{k(q_{\alpha}-1)}{2}\mod -1, \quad h_{\alpha}=\tilde{h}_{\alpha}-\frac{1}{2}=\frac{q_{\alpha}}{2}.
  \end{equation}
The vacuum $|k\rangle$ is annihilated by the positive modes and by $\chi_0$ in case such a fermionic zero mode exists. The OPEs (\ref{freeope}) encode the following (anti-)commutators:
\begin{equation}
  [y_r^{\alpha},\rho_{\beta s}]=\delta^{\alpha}_{\beta}\delta_{r,-s}, \qquad [\chi_r^{\alpha},\overline{\chi}_{\beta s}]=\delta^{\alpha}_{\beta}\delta_{r,-s}. 
\end{equation}
The left and right R-charges of $|k\rangle$ have been computed in \cite{Bertolini:2013xga}:
\begin{align}
  \label{vacrcharges}
  q_{|k\rangle}&=\sum_{\alpha}\left[(q_{\alpha}-1)(\tilde{\nu}_{\alpha}-\frac{1}{2})-q_{\alpha}(\nu_{\alpha}-\frac{1}{2}) \right],\nonumber\\
  \overline{q}_{|k\rangle}&=\sum_{\alpha}\left[q_{\alpha}(\tilde{\nu}_{\alpha}+\frac{1}{2})+(q_{\alpha}-1)(-\nu_{\alpha}+\frac{1}{2}) \right].
  \end{align}
Furthermore, $E_{|k\rangle}=0$ for even $k$ and
\begin{equation}
  E_{|k\rangle}=-\frac{5}{8}+\frac{1}{2}\sum_{\alpha}[\nu_{\alpha}(1-\nu_{\alpha})+\tilde{\nu}_{\alpha}(1+\tilde{\nu}_{\alpha})]
  \end{equation}
for $k$ odd. Using $K_Y=\mathcal{O}_Y$ and $X=\oplus_i L_i$, the state $|k\rangle$ transforms as a section of $L_{|k\rangle}$ with
\begin{equation}
  L_{|k\rangle}=\left\{
\begin{array}{ll}
  \otimes_iL_i^{\tilde{\nu}_i-\nu_i} & k \textrm{ even},\\
  \otimes_iL_i^{\tilde{\nu}_i-\nu_i+\frac{1}{2}} & k \textrm{ odd.}
  \end{array}\right.
\end{equation}
Since $q_I=0$ for the base coordinates, we have $\nu_I=0$ in all sectors and $\tilde{\nu}_I=-\frac{k}{2}\mod -1$. Defining $\tau_{\alpha}=\nu_{\alpha}-\tilde{\nu}_{\alpha}$, one easily confirms
\begin{equation}
  \label{taudef}
  \tau_{\alpha}=\left\{\begin{array}{ll}
  0&\nu_{\alpha}=0\\
  1&\nu_{\alpha}\neq 0
\end{array}
  \right. k\textrm{ even};\qquad
\tau_{\alpha}=\left\{\begin{array}{ll}
  \frac{1}{2}&\nu_{\alpha}\leq\frac{1}{2}\\
  \frac{3}{2}&\nu_{\alpha}>\frac{1}{2}
\end{array}
  \right. k\textrm{ odd}.
  \end{equation}
Therefore, ground states are of the form
\begin{equation}
  |\Psi^k_0\rangle=\Psi_0(y',\overline{y})_{\overline{I}_1,\ldots,\overline{I}_u}\overline{\eta}^{\overline{I}_1}\cdot\ldots\cdot\overline{\eta}^{\overline{I}_u}|k\rangle.
\end{equation}
Here, the $y'$ denote zero modes. The $\Psi^k$ are $(0,u)$-forms on $Y$ taking values in $L^{\ast}_{|k\rangle}$. The state above is a generic ground state. There can be further ground states in sectors that have additional zero modes. 

Once we have constructed a Hilbert space of states, we can compute the $\overline{Q}$-cohomology. We take the restricted Hilbert space $\mathcal{H}$ with fixed $E$ and $q$. This is further graded by the $U(1)_R$-charge and $\overline{Q}:\mathcal{H}_q\rightarrow\mathcal{H}_{q+1}$. The key to computing the cohomology is that $\overline{Q}=\overline{Q}_0+\overline{Q}_W$ generates a double-complex. To see this, one introduces an operator $U$ under which $\overline{\eta}$ and $\eta$ are charged by $+1$ and $-1$, respectively, and the other fields are invariant. We denote the eigenvalues of $U$ by $u$ and define $p=q-u$. One can show that $U$ commutes with $\overline{Q}_0$ and $\overline{Q}_W$. Hence, one obtains a double complex with differentials $\overline{Q}_0$ and $\overline{Q}_W$ acting as
\begin{equation}
  \overline{Q}_0:\mathcal{H}^{p,u}\rightarrow \mathcal{H}^{p,u+1}, \qquad \overline{Q}_W:\mathcal{H}^{p,u}\rightarrow\mathcal{H}^{p+1,u}.
\end{equation}
In other words, the cohomology splits with respect to the base and the fibre direction, where $\overline{Q}_0$ acts horizontally along the base, while $\overline{Q}_W$ acts in the fibre direction. We thus obtain a spectral sequence with
\begin{equation}
  E_1^{p,u}=H_{\overline{Q}_0}^u(\mathcal{H}^{p,\bullet}), \qquad\qquad E_2^{p,u}=H^p_{\overline{Q}_W}H_{\overline{Q}_0}^u(\mathcal{H}^{\bullet,\bullet}).
\end{equation}
The differentials are $d_0=\overline{Q}_0$, $d_1$, and $d_r:E_r^{p,u}\rightarrow E_r^{p+r,u+1-r}$ with $d_r$ vanishing for $r>\mathrm{dim}B$. The spectral sequence converges:
\begin{equation}
  E_{\mathrm{dim}B+1}^{p,u}=E_{\infty}^{p,u}=H_{\overline{Q}}^{p,u}(\mathcal{H}^{\bullet,\bullet}).
  \end{equation}
Using this spectral sequence, one can compute the cohomology by first computing the cohomology of
\begin{equation}
  \overline{Q}_0=-\overline{\eta}^{\overline{I}}\frac{\partial}{\partial\overline{y}^{\overline{I}}}
  \end{equation}
on $\mathcal{H}$ and then computing the $\overline{Q}_W$ cohomology.

Similar to what is known from Landau-Ginzburg orbifolds \cite{Vafa:1989xc,Intriligator:1990ua}, we have to project onto invariant states in each twisted sector. For hybrids, the new feature is that fibre coordinates may also transform non-trivially with respect to the base, so only states with certain times of insertions of fields living on the base will survive the projection.
%%%%%%
\subsubsection{(R,R)-sectors}
The situation simplifies when we focus on (R,R)-sectors. For these sectors, $E_{|k\rangle}=0$ and one can restrict to zero-modes on all fields. There is a further simplification for those twisted sectors where the Landau-Ginzburg fibre does not contribute any light degrees of freedom. In this case the cohomology of $\overline{Q}_W$ is trivial and the massless spectrum is governed by the cohomology of the base. As we will show, this is exactly what determines the narrow sectors.

Following \cite{Bertolini:2013xga}, we organise the $y^{\alpha}$ into light and heavy fields. This is determined by $\tau_{\alpha}$ in (\ref{taudef}). We perform a split $y^{\alpha}=(y^{\alpha'},\phi^A)$ with $\tau_{\alpha'}<1$ (``light'') and $\tau_A\geq 1$ (``heavy''). Since all the base fields have $\tau_I<1$, we further have $y^{\alpha'}=(y^I,\phi^{i'})$. Then the bundle $X$ can be decomposed as $X=X_k\oplus\oplus_AL_A$. Furthermore, one defines $Y_k=\mathrm{tot}(X_k\stackrel{\pi_k}{\rightarrow}B)$. It is easy to see from (\ref{modes}) that only the light fields have zero modes. Thus, a general state has the form
\begin{equation}
  \label{genstate}
  |\Psi_u^s\rangle=\Psi(y',\overline{x})^{\alpha'_1\ldots\alpha'_s}_{\overline{I}_1\ldots\overline{I}_u}\overline{\chi}_{\alpha_1'}\cdot\ldots\cdot\overline{\chi}_{\alpha_s'}\overline{\eta}^{\overline{I}_1}\cdot\ldots\cdot\overline{\eta}^{\overline{I}_u}|k\rangle.
\end{equation}
Geometrically, $\overline{\chi}_{\alpha'}$ transforms as a section of $T^{\ast}_{Y_k}$ and $\overline{\eta}^{\overline{I}}$ transforms as a section of $\pi^{\ast}_k(\overline{T}_B)$. The vacuum $|k\rangle$ is a section of $L_{|k\rangle}=\pi^{\ast}_k(\otimes_AL^{\ast}_A)$. It follows that $\Psi_u^s$ is a $(0,u)$-form valued in $\mathcal{E}^s=\wedge^sT_{Y_k}\otimes L^{\ast}_{|k\rangle}$.

Restricting the Killing vector field $V$ to $Y_k$, one has $\mathcal{L}_V\Psi=q_{\Psi}\Psi$. From this, one obtains the following left and right R-charges of $|\Psi\rangle$:
\begin{equation}
  q=q_{|k\rangle}+q_{\Psi}+s,\qquad\qquad \overline{q}=\overline{q}_{|k\rangle}+q_{\Psi}+u. 
\end{equation}
Acting with $\overline{Q}_0$ on $|\Psi_s^u\rangle$ gives $\Psi_s^u\rightarrow -\overline{\partial}\Psi_{u+1}^s$, so that one has to compute the cohomology $H^{\bullet}_r(Y_k,\mathcal{E}^{\bullet})$, where the grading $r=(r_1,\ldots,r_n)$ is determined by monomials $\prod_i\phi_i^{r_i}$ in the fibre fields that arise in generic states. The operator $\overline{Q}_W$ has the form
\begin{equation}
  \overline{Q}_W=W_{\alpha'}(y')\chi^{\alpha'}
\end{equation}
and acts on the state as
\begin{equation}
  \overline{Q}_W:\Psi_u^s\mapsto(sW_{\alpha'}\Psi^{\alpha_1'\ldots\alpha_s'})_u^{s-1}. 
\end{equation}
We will see in our examples that, when the set of fields $y'$ is empty, the operator is zero because $W_{\alpha'}(y')=0$. It would be interesting to see if this holds in general for good hybrids.

Finally, note that \cite{Bertolini:2013xga} also discusses a pairing on the state space. CPT invariance implies that for each state with charge $(q,\overline{q})$ in the $k$-sector there should be a corresponding state state with $(-q,-\overline{q})$ in the $2N-k$ sector. Under the assumption that all fields have $\tilde{\nu}<0$, the authors argue that CPT invariance reduces to Serre duality for the cohomology on $B$. While this assumption is not satisfied for generic (R,R)-sectors, the condition always seems to be satisfied for the narrow sectors discussed below. Below, we will propose a pairing on the narrow state space that corresponds to the topological pairing rather than the CPT pairing. 
%%%%%
\subsubsection{Narrow sectors, chiral rings and marginal deformations}
\label{sec-hybnarrow}
For our purposes we do not need the full spectrum of ground states of the theory but only specific ones that are relevant for understanding the $I$- and $J$-functions. The relevant states are elements in the $(a,c)$-ring. The $(a,c)$ states in turn are in one-to-one correspondence with states in the (R,R)-sector, so we can restrict to twisted sectors with $k\in2\mathbb{Z}_{\geq 0}$ from now on.

To make the connection between the ground states in the (R,R)-sector and the relevant elements of the chiral ring, we apply spectral flow \cite{Lerche:1989uy}. Let $\mathcal{U}_{\theta,\overline{\theta}}$ denote the spectral flow operator. Given a CFT with central charge $c=3\widehat{c}$, the left and right R-charges get shifted by $(\widehat{c}\theta,\widehat{c}\overline{\theta})$.  To get to the elements of the $(a,c)$-ring we have to choose $(\theta,\overline{\theta})=(-\frac{1}{2},\frac{1}{2})$. The corresponding $(a,c)$-ring elements will then have $0\leq \overline{q}\leq\widehat{c}$ and $0\leq-q\leq\widehat{c}$. Among these states there is a unique $(a,c)$-ring element with $(q,\overline{q})=(0,0)$ (corresponding to the (R,R)-ground state with $(q,\overline{q})=(\frac{3}{2},-\frac{3}{2})$) which we identify with the vacuum. There is also a unique state with $(q,\overline{q})=(-\widehat{c},\widehat{c})$. A further subset of the $(a,c)$-ring elements are those that correspond to a subset of marginal deformations of the Calabi-Yau CFT. These are the states satisfying $(q,\overline{q})=(-1,1)$. In the context of FJRW theory, we have to restrict to narrow states, which in general only account for a subring of the $(a,c)$-ring. We claim that the narrow sector states correspond to states in those twisted sectors with even $k$ where all fibre fields $\phi^i$ are heavy, i.e.~they have $\tau_i\geq 1$ and hence do not contribute any zero modes. This has several consequences:
\begin{itemize}
\item Since $X_k$ is trivial, $Y_k=B$. Hence $\overline{\chi}_{\alpha'}\equiv\overline{\chi}_I$ are sections of $T^*_B$. The wave functions $\Psi^s_u$ in the narrow sectors are horizontal $(0,u)$-forms taking values in $\wedge^sT_B\otimes L^{\ast}_{|k\rangle}$. Furthermore recall that the $\overline{\eta}^{\overline{I}}$ are sections of $\overline{T}_B$.
\item The vacuum $|k\rangle$ is a section of $L_{|k\rangle}=\pi_k^{*}(\otimes_iL_i^*)$ where $i$ accounts for all the fibre fields. In other words, $\{\phi^A\}=\{\phi^i\}$. By construction, this is invariant under the action of the orbifold group and hence $|k\rangle$ does not get projected out. It is in fact the only state in the fibre direction.
  \item The bundle $L_{|k\rangle}$ transforms as a section of a bundle over $B$. To obtain an invariant state, this has to be compensated by taking $|\Psi_s^u\rangle$ to have a suitable number of insertions of $\overline{\chi}_I$ and $\overline{\eta}^{\overline{I}}$. 
  \item Since for all the models we consider $W_{\alpha'}(y')=0$ if all fibre fields are heavy, $\overline{Q}_W=0$ in the narrow sectors and the cohomology of $\overline{Q}$ is equivalent to the cohomology of $\overline{Q}_0$. 
  \end{itemize}
Modulo subtleties related to gerbes, $B=\mathbb{P}^n$ in all our examples. Hence all the $\overline{\chi}_I$ and $\overline{\eta}^{\overline{I}}$ are equivalent in cohomology. Therefore every $|\Psi_u^s\rangle$ for fixed $s,u$ corresponds to the same element in cohomology. 

Since spectral flow provides a one-to-one correspondence between (R,R) ground states and $(a,c)$-ring elements, we can indirectly characterise elements of the narrow $(a,c)$-ring by the corresponding (R,R)-states. This is enough for our purposes, as we do not require explicit representatives of the $(a,c)$-states. We denote the relevant state space by 
\begin{equation}
\mathcal{H}^{(a,c)}_{\mathrm{nar}}=\bigoplus_{\delta}H_{\delta}^{\ast}(B)=\bigoplus_{\delta,a}\mathcal{H}^{(a,c)}_{\delta,a}.
\end{equation}
Disentangling the $\mathbb{Z}_2$ from the GSO projection and the action of the orbifold group, $\delta=\frac{k}{2}$ labels the contributing $\delta$-twisted sector of the $\mathbb{Z}_N$-orbifold and $a$ runs over the cohomology elements of $H^{\ast}(B)$. For example, for $B=\mathbb{P}^2$ we have $a\in\{{\bf 1},H,H^2\}$ where $H$ is the hyperplane class of $\mathbb{P}^2$. We get a copy of $H^{\ast}(B)$ for each narrow twisted sector. The spaces $\mathcal{H}^{(a,c)}_{\delta,a}$ are one dimensional and we denote the corresponding basis elements by $e_{\delta,a}$. Given this, we propose the following pairing on the narrow state space: 
\begin{equation}
  \label{pairing}
  \langle e_{i,a} , e_{j,b}\rangle=\frac{1}{N}\delta_{i,N-j}\int_Ba\wedge b.
\end{equation}
In the special cases of $\mathbb{Z}_N$-Landau-Ginzburg orbifolds and Calabi-Yau complete intersections in a toric ambient space, this reduces to the well-known pairings, corresponding to the topological pairing on the worldsheet CFT. In particular, this pairing determines the components of the topological metric.

Once we have computed the $I$-function $I_{\delta}(u(\mathtt{t}),H)$ from the GLSM, we can expand it in terms of $H$ and then match the components labelled by $\delta$ and the degree of $H$ with elements of the state space of the hybrid theory. To match with the labels of the twisted sectors as they arise from the GLSM we have $\delta=\frac{k}{2}$. In concrete examples, it is straightforward to match the components of the $I$-function with the respective states of the hybrid theory. Identifying the distinguished $(a,c)$-ring elements with left and right R-charges $(0,0)$ and $(-1,1)$ we can extract the coordinate transformation to compute the $J$-function from the $I$-function. 
%%%%%%%%%%%%%%%%%%%%%%%%%%%%%%%%%%%%%%%%%%%%%%%%%%%%%%%%%%%%%%%%%%%%%%%%%%%%%
\section{One-parameter examples}
\label{sec-1par}
In this section, we compute invariants of hybrid phases of a class of one-parameter models. The corresponding GLSMs and hybrid phases have been described at length in \cite{Erkinger:2020cjt} and the $I$-functions were extracted from the sphere partition function and shown to match with results in the mathematics literature where available. We use the same labelling for the models as in \cite{Erkinger:2020cjt}.

\subsection{Model K1}
\label{sec-k1}
The GLSM associated to this well-studied model has gauge group $\mathsf{G}=U(1)$ with the following matter content
\begin{equation}
  \begin{array}{c|cc|c}
    &p_1,p_2&x_1,\ldots,x_6&\mathrm{FI}\\
    \hline
    U(1)&-3&1&\zeta\\
    U(1)_V&2-6q&2q&
    \end{array}
\end{equation}
Here, $U(1)_V$ is the vector $U(1)$, $\mathrm{FI}$ denotes the FI-parameter of the GLSM, and $0\leq q\leq\frac{1}{3}$. The GLSM superpotential is $W=p_1G^1_3(x_1,\ldots,x_6)+p_2G_3^2(x_1,\ldots,x_6)$, where $G_3^1,G_3^2$ are suitably generic homogeneous polynomials of degree $3$. In the $\zeta\gg0$-phase, one has to set $q=0$ to match with the R-charges of the low-energy nonlinear sigma model and one recovers the Calabi-Yau $\mathbb{P}^5[3,3]$, i.e.~a codimension $2$ complete intersection of two cubics $\{G_3^1=0,G_3^2=0\}$ in $\mathbb{P}^5$.

We are interested in the hybrid phase at $\zeta\ll0$. In this phase, the ground state is given by $x_i=0$. The D-term equation of the GLSM reduces to
\begin{equation}
  -3|p_1|^2-3|p_2|^2=\zeta\ll 0.
\end{equation}
The gauge symmetry is broken to $G=\mathbb{Z}_3$. To understand this phase, it makes sense to reverse the $U(1)$-charges of the fields and the sign of $\zeta$ so that $\zeta$ can be interpreted as a volume parameter. Then the $p$-fields take values in $B=\mathbb{P}^1_{33}$, i.e.~a $\mathbb{P}^1$ whose homogeneous coordinates have weight $3$. Taking into account classical fluctuations of the $x_i$, one finds that over each point $(\langle p_1\rangle,\langle p_2\rangle)$ there is a Landau-Ginzburg orbifold with superpotential  $W=\langle p_1\rangle G^1_3(x_1,\ldots,x_6)+\langle p_2\rangle G_3^2(x_1,\ldots,x_6)$. The $\mathbb{Z}_3$-orbifold group comes from the broken $U(1)$ of the GLSM. The R-charges of the matter fields in the hybrid theory are obtained by choosing $q=\frac{1}{3}$ which implies that, in the notation of Section~\ref{sec-hyb}, the base coordinates $y^I=\{p_1,p_2\}$ have $(q,\overline{q})=(0,0)$ and the fibre coordinates $\phi^i=\{x_1,\ldots,x_6\}$ have $(q,\overline{q})=(\frac{1}{3},\frac{1}{3})$.

This model, along with all other one-parameter models we discuss, is an example of a hybrid where the orbifold group of the fibered Landau-Ginzburg model also acts on the base, albeit trivially. The hybrid model thus has a gerbe structure. This has already been observed in \cite{Caldararu:2007tc}, where this configuration was denoted as $\mathcal{O}(-\frac{1}{3})^{\oplus 6}\rightarrow G_3\mathbb{P}^1$, with $G_3\mathbb{P}^1$ a $\mathbb{Z}_3$-gerbe. It was then argued, in line with the GLSMs and monodromy considerations, that the hybrid model can be described as $\mathcal{O}(-1)^{\oplus 6}\rightarrow\mathbb{P}^1$, i.e.~ $B=\mathbb{P}^1$ and $X=\mathcal{O}(-1)^{\oplus 6}$ in the notation of \cite{Bertolini:2013xga}. This is also consistent with FJRW theory \cite{MR3590512}. For the sake of constructing the states in the conformal field theory we can assume that $B=\mathbb{P}^1$ because the R-charges of the base coordinates are zero. However, the fact that we actually have $B=\mathbb{P}^1_{33}$ is important for the selection rule (\ref{rule2}) and the projection onto invariant states.

\subsubsection{State space and narrow sectors}
Next, we analyse the state space in the hybrid model and identify the narrow sectors. Inserting into (\ref{vacrcharges}) we find the following charges $(q,\overline{q})$ of the vacua $|k\rangle$ of the $k$-th twisted sector of the $\mathbb{Z}_6$-orbifold that combines the orbifold $\mathbb{Z}_3$ and the GSO $\mathbb{Z}_2$:
\begingroup
\renewcommand{\arraystretch}{1.2}
\begin{equation}
  \begin{array}{c|rrrrrr}
    k&0&1&2&3&4&5\\
    \hline
    q&-\frac{3}{2}&0&\frac{1}{2}&-2&-\frac{3}{2}&0\\
    \hline
    \overline{q}&-\frac{3}{2}&-\frac{3}{2}&-\frac{3}{2}&\frac{1}{2}&\frac{1}{2}&\frac{1}{2}
    \end{array}
\end{equation}
\endgroup
Since all the fibre fields have $q^i=\frac{1}{3}$, we compute the $\nu_i$, $\tilde{\nu}_i$ and $\tau_i$, defined in (\ref{nudef}) and (\ref{taudef}), for each twisted sector:
\begingroup
\renewcommand{\arraystretch}{1.2}
\begin{equation}
  \begin{array}{c|rrrrrr}
    k&0&1&2&3&4&5\\
    \hline
    \nu_i&0&\frac{1}{6}&\frac{1}{3}&\frac{1}{2}&\frac{2}{3}&\frac{5}{6}\\
    \hline
    \tilde{\nu}_i&0&-\frac{1}{3}&-\frac{2}{3}&0&-\frac{1}{3}&-\frac{2}{3}\\
    \hline
    \tau_i&0&\frac{1}{2}&1&\frac{1}{2}&1&\frac{3}{2}
    \end{array}
\end{equation}
\endgroup
From this we can read off that in the sectors $k=2$ and $k=4$ all the fields are heavy, i.e.~$\{\phi^i\}=\{\phi^A\}$ and $y^{\alpha'}=\{p_1,p_2\}$. The states in these sectors are thus determined by the base degrees of freedom alone. The fibre component of the state is the vacuum $|k\rangle$ for $k=2,4$ which transforms as $L_{|k\rangle}=\otimes_i\mathcal{O}(-1)^{-1}=\mathcal{O}(6)$. This is the same as the anti-canonical bundle of $B$, due to the non-trivial weights of the base coordinates. The vacua $|k\rangle$ in these sectors are invariant under the $\mathbb{Z}_3$-orbifold. The fermions $\overline{\chi}_I$ and $\overline{\eta}^{\overline I}$, with $I=1$ are sections of $T^*_B$ and $\overline{T}_B$ which is $\mathcal{O}(-6)$ in both cases. 

In a local patch on $B$ we can construct the following states $|\Psi_u^s\rangle_{q,\overline{q}}$ is the narrow twisted sectors:
\begin{align}
  \label{33states}
  k&=2:&|\Psi_{0}^0\rangle_{\frac{1}{2},-\frac{3}{2}}&&|\Psi_{0}^1\rangle_{\frac{3}{2},-\frac{3}{2}}&&|\Psi_{1}^0\rangle_{\frac{1}{2},-\frac{1}{2}}&&|\Psi_{1}^1\rangle_{\frac{3}{2},-\frac{1}{2}}\nonumber\\
  k&=4:& |\Psi_{0}^0\rangle_{-\frac{3}{2},\frac{1}{2}}&&|\Psi_{0}^1\rangle_{-\frac{1}{2},\frac{1}{2}}&&|\Psi_{1}^0\rangle_{-\frac{3}{2},\frac{3}{2}}&&|\Psi_{1}^1\rangle_{-\frac{1}{2},\frac{3}{2}}
\end{align}
Due to the transformation properties of $|k\rangle$, we have to insert either a $\overline{\chi}_I$ or an $\overline{\eta}^{\overline I}$ to get an invariant state. This removes all the states with $|q|\neq|\overline{q}|$, as expected. These states are in the $\overline{Q}_0$-cohomology. Since $W_{\alpha'}(y')=0$ all states are automatically in the $\overline{Q}_W$-cohomology.

In summary, the state space consists of
\begin{align}
  \label{33statesred}
  k&=2:&|\Psi_{0}^1\rangle_{\frac{3}{2},-\frac{3}{2}}&&|\Psi_{1}^0\rangle_{\frac{1}{2},-\frac{1}{2}}\nonumber\\
  k&=4:&|\Psi_{0}^1\rangle_{-\frac{1}{2},\frac{1}{2}}&&|\Psi_{1}^0\rangle_{-\frac{3}{2},\frac{3}{2}}
\end{align}

Applying spectral flow to get to the $(a,c)$-ring shifts the R-charges by $(-\frac{3}{2},\frac{3}{2})$. The state $|\Psi_{0}^1\rangle_{\frac{3}{2},-\frac{3}{2}}$ at $k=2$ maps to an $(a,c)$-ring element with $(q,\overline{q})=(0,0)$ and the state $|\Psi_{1}^0\rangle_{\frac{1}{2},-\frac{1}{2}}$ corresponds to the $(a,c)$-ring element with  $(q,\overline{q})=(-1,1)$. They represent the cohomology classes of $H^*(\mathbb{P}^1)$ whose basis elements we denote by $e_{1,{\bf 1}}$ and $e_{1,H}$, where $H$ is the hyperplane class of $\mathbb{P}^1$ and the labelling anticipates that we will match $\delta=\frac{k}{2}$ for the $\delta$-twisted sector of the $\mathbb{Z}_3$-orbifold. We get another copy of $H^*(\mathbb{P}^1)$ from the sector $k=4$. We will denote the corresponding basis elements by  $e_{2,{\bf 1}}$ and $e_{2,H}$.  Using the correspondence between the sectors $k$ and $6-k$, we can define a pairing 
\begin{equation}
  \label{k1pairing}
  \langle e_{i,a},e_{j,b}\rangle=\frac{1}{3}\delta_{i,3-j}\int_{\mathbb{P}^1}a\wedge b.
\end{equation}
Hence we have $\langle e_{1,{\bf 1}},e_{2,H}\rangle=\langle e_{1,H},e_{2,{\bf 1}}\rangle=\frac{1}{3}$ and zero for all other pairings. 
%%%%%
\subsubsection{$J$-function and invariants}
In \cite{Erkinger:2020cjt}, the $I$-function of this model was extracted from the GLSM sphere partition function and confirmed to coincide with the result of \cite{MR3590512}. From the expression (\ref{hybifun}), the following result for $I_{\delta}(\mathsf{t},H)$ was found for this model:
\begin{equation}
  \label{k1ifun}
  I_{\delta}^{\zeta\ll 0}(\mathsf{t},H)=\frac{\Gamma\left(1+\frac{H}{2\pi i}\right)^2}{\Gamma\left(\frac{H}{6\pi i}+\left\langle\frac{\delta}{3}\right\rangle\right)^6}\sum_{n=0}^{\infty}\frac{\Gamma\left(n+\frac{H}{6\pi i}+\frac{\delta}{3}\right)^6}{\Gamma\left(3n+\delta+\frac{H}{2\pi i}\right)^2}e^{\mathsf{t}\left(\frac{H}{6\pi i}+n+\frac{\delta}{3}-\frac{1}{3}\right)},
\end{equation}
where $\mathsf{t}=2\pi \zeta-i\theta$ with theta angle $\theta$, $\langle x\rangle=x-\lfloor x\rfloor$, and $\delta=1,2$. Expanding this further up to order $H$, noting that $\int_{\mathbb{P}^1}H^2=0$, we denote the respective terms in the expansion by $I_{\delta}^{\bf 1}(\mathsf{t})$ and $I_{\delta}^{H}(\mathsf{t})$. Then we can write the $I$-function in terms of the basis of $\mathcal{H}^{(a,c)}_{\mathrm{nar}}$ as
\begin{equation}
  I(\mathsf{t})=I_{1}^{\bf 1}e_{1,{\bf 1}}+I_{1}^{H}e_{1,H}+I_{2}^{\bf 1}e_{2,{\bf 1}}+I_{2}^{H}e_{2,H}.
\end{equation}
From this we read off the coordinate transformation that will give us the $J$-function:
\begin{equation}
  t(\mathsf{t})=\frac{I_{1}^{H}(u(\mathsf{t}))}{I_{1}^{\bf 1}(u(\mathsf{t}))}.
\end{equation}
where we set $u=e^{\frac{\mathsf{t}}{3}}$. We insert this into the definition of the $J$-function and compare with the general form (\ref{hyb-jfunction}). By (\ref{rule1}), we can only have the following correlators:
\begin{equation}
  \label{k1rule1}
  \langle \tau_1(e_{1,1})(e_{1,H})^k\rangle_{0,k+1,\beta}, \qquad \langle (e_{1,H})^{k+1}\rangle_{0,k+1,\beta}.
\end{equation}
For the first correlator we can use the dilaton equation (\ref{dilatoneq}) to rewrite it as
\begin{equation}
  \langle \tau_1(e_{1,1})(e_{1,H})^k\rangle_{0,k+1,\beta}=(-2+k)\langle(e_{1,H})^k\rangle_{0,k,\beta}.
  \end{equation}
Moreover, making use of the divisor equation (\ref{divisoreq}), we can write
\begin{equation}
  \langle(e_{1,H})^k\rangle_{0,k,\beta}=\left(\int_{\beta}e_{1,H}\right)^k\langle \cdot \rangle_{0,0,\beta}=\beta^k \langle \cdot \rangle_{0,0,\beta}.
\end{equation}
Finally, we can apply the selection rule (\ref{rule2}). For this model, this has been stated in \cite{MR3590512} for a correlator with $m$ insertions and an orbifold group $\mathbb{Z}_N$:
\begin{equation}
  \label{rule2clader}
  2g-2+m-\beta-\sum_{i=1}^m\delta_{j_i}=0\mod N.
  \end{equation}
The base coordinates also transform under the quantum symmetry group which makes the appearance of a term involving $\beta$ plausible from the physics perspective.

Applying this selection rule, we get for both types of correlators
\begin{equation}
  -2-\beta=0\mod 3\qquad\leftrightarrow \qquad \beta=3n-2, \quad n=1,2,\ldots
\end{equation}
Finally, we make use of the pairing (\ref{k1pairing}) to deduce the following structure of the $J$-function:
\begin{align}
  \label{k1jfun}
  J=&e_{1,{\bf 1}}+te_{1,H}+\sum_{k\geq0}\sum_{n\geq 1}\frac{t^k}{k!}\left(\int_{[3n-2]}e_{1,H}\right)^{k+1}\langle\cdot\rangle_{0,0,3n-2}3e_{2,{\bf 1}}\nonumber\\
  &\quad\quad\quad\quad\quad+\sum_{k\geq0}\sum_{n\geq 1}\frac{t^k}{k!}(-2+k)\left(\int_{[3n-2]}e_{1,H}\right)^{k}\langle\cdot\rangle_{0,0,3n-2}3e_{2,H}\nonumber\\
  =&e_{1,{\bf 1}}+te_{1,H}+3e_{2,{\bf 1}}\sum_{n\geq 1}e^{(3n-2)t}(3n-2)\langle\cdot\rangle_{0,0,3n-2}\nonumber\\
  &\qquad\qquad\quad+3e_{2,H}\sum_{n\geq 1}e^{(3n-2)t}[(3n-2)t-2]\langle\cdot\rangle_{0,0,3n-2}
\end{align}
Inserting the $I$-function (\ref{k1ifun}) into the definition (\ref{ijrel}) of the $J$-function, we can use the this expression to read off the invariants. The first step is the coordinate transformation. Absorbing the factor $2\pi i$ into the definition of $H$,  $\frac{H}{2\pi i}\rightarrow H$, we find
\begin{equation}
  t=\frac{I_1^H(u)}{I_1^{\bf 1}(u)}=\log u+\frac{7 u^3}{78732}+\frac{9109 u^6}{258280326000}+\frac{4178309u^9}{183014339639688000}+O\left(u^{11}\right).
\end{equation}
Defining $q=e^t$ and exponentiating and inverting the series we get
\begin{equation}
  \label{k1mir}
  u(q)=q-\frac{7 q^4}{78732}-\frac{11779 q^7}{1549681956000}-\frac{21607
   q^{10}}{6778308875544000}+O\left(q^{12}\right).
\end{equation}
To extract the invariants, we insert this into $\frac{I_2^{\bf 1}}{I_1^{\bf 1}}$:
\begin{equation}
\frac{I_2^{\bf 1}}{I_1^{\bf 1}}=q+\frac{q^4}{39366}+\frac{501119 q^7}{75934415844000}+\frac{139
   q^{10}}{50209695374400}+\frac{1568601780510977
   q^{13}}{1080388389410293928915808000000}+O\left(q^{16}\right)
\end{equation}
Comparing with the $e_{2,{\bf 1}}$-coefficient in (\ref{k1jfun}) and replacing $t=\log q$ yields the following non-zero correlators:
\begingroup
\renewcommand{\arraystretch}{1.3}
\begin{equation}
  \begin{array}{c|c|c|c|c|c}
    3n-2&1&4&7&10&13\\
    \hline
    \langle\cdot\rangle_{3n-2}&\frac{1}{3}&\frac{1}{472392}&\frac{501119}{1594622732724000}&\frac{139}{1506290861232000}&\frac{1568601780510977}{42135147187001463227716512000000}
    \end{array}
\end{equation}
\endgroup
The information from $\frac{I_2^{H}}{I_1^{\bf 1}}$ is redundant and gives the same invariants. 
%%%%%
\subsubsection{Compatibility with mirror symmetry}
\label{sec-mir33}
To check the consistency of our results, we can make use of the fact that the mirrors in the geometric phase of our models are well-studied. Using standard techniques\footnote{See for instance \cite{Cox:2000vi} for a standard reference.}, one can compute the Yukawa couplings in the large complex structure limit of the mirror B-model. The B-model Yukawa couplings in the hybrid phase are then obtained by a simple change of coordinates. Adapting a discussion outlined in \cite{Huang:2006hq} for Landau-Ginzburg regions of the moduli space, we reinterpret (\ref{k1mir}) as the mirror map, insert it into the Yukawa couplings, and compare the result with the $J$-function. Using $N=2$ special geometry for $\widehat{c}=3$, it was argued in \cite{Knapp:2020oba} that the $J$-function can be expressed as
\begin{equation}
  \label{jprepot}
  J=e_0+\sum_{i=1}^h\left(t_ie_i+\frac{\partial\mathcal{F}}{\partial t_i}e^i\right)+\left(\frac{\partial\mathcal{F}}{\partial t_i}-2\mathcal{F}\right)e^0,
\end{equation}
where $\mathcal{F}$ is the prepotential from which one obtains the Yukawa couplings as
\begin{equation}
  C_{ijk}=\frac{\partial^3\mathcal{F}}{\partial t_i\partial t_j\partial t_k}.
\end{equation}
An alternative way to compute derivatives of the components of the $J$-function is thus via the Yukawa couplings of the mirror.

Let us apply this to the present example. The Picard-Fuchs operator in the large complex structure limit of the mirror is
\begin{equation}
  \mathcal{L}=\theta^4-9z(3\theta+1)^2(3\theta+2)^2,
\end{equation}
where $\theta=z\frac{d}{dz}$ and we can identify $z=e^{-\mathsf{t}}=u^{-3}$. Recall that the B-model Yukawa coupling is defined as
\begin{equation}
  Y_{zzz}=\int_{\widetilde{X}}\Omega\wedge\Omega''', \qquad \Omega'=\theta\Omega,
\end{equation}
where $\widetilde{X}$ is the mirror Calabi-Yau and $\Omega(z)$ is the holomorphic threeform. Using the Picard-Fuchs equation, one can derive a first order differential equation for $Y_{zzz}$. Furthermore, the mirror map is defined by $t(z)=\frac{\varpi_1(z)}{\varpi_0(z)}$ where $\varpi_0,\varpi_1$ are solutions to the Picard-Fuchs equation such that $\varpi_0=1+\ldots$ and\footnote{To be precise, monodromy dictates that that $\varpi_1$ is normalised as $\varpi_1=\frac{1}{2\pi i}\log z+\ldots$. We sometimes suppress these factors for ease of notation.} $\varpi_1=\log z+\ldots$. Recalling that we defined $q=e^{t}$ and $\frac{d}{dq}=\frac{dz}{dq}\frac{d}{dz}$ and normalising the Yukawa coupling by $\frac{1}{\varpi_0^2}$, the A-model and B-model Yukawa couplings are related as follows:
\begin{equation}
  C_{ttt}(q)=\frac{1}{\varpi_0^2(z(q))}\left(\frac{q}{z}\frac{dz}{dq}\right)^3Y_{zzz}(z(q)).
\end{equation}
In the present example, the B-model Yukawa coupling is
\begin{equation}
  Y_{zzz}=\frac{9}{1-729 z}.
\end{equation}
We have chosen the normalisation to match with the triple intersection number of $\mathbb{P}^5[3,3]$. Rescaling the holomorphic three-form\footnote{In GLSM language, this accounts for the relative shift in R-charges between the large volume and the hybrid phase.} $\Omega(z)\rightarrow u\Omega(u)$, the Yukawa coupling in the $u$-coordinates is
\begin{equation}
  Y_{uuu}=\frac{243u}{729-u^3}. 
\end{equation}
Further implementing the coordinate change (\ref{k1mir}), now interpreted as the mirror map, we get
\begin{equation}
  C_{ttt}=\frac{q}{3}+\frac{8 q^4}{59049}+\frac{501119 q^7}{4649045868000}+\frac{139q^{10}}{1506290861232}%+\frac{1568601780510977q^{13}}{19178492119709359684896000000}
  +O\left(q^{11}\right).
\end{equation}
Recalling that, by (\ref{k1pairing}), the dual of $e_{1,H}$ is $3e_{2,{\bf 1}}$, one, confirms that
\begin{equation}
  C_{ttt}=\theta_q^2\left(\frac{1}{3}\frac{I_2^{\bf 1}(q)}{I_1^{\bf 1}(q)}\right), \qquad\qquad \theta_q=q\frac{d}{dq}.
\end{equation}
This is consistent with (\ref{jprepot}) and also shows that the normalisation of the invariants is compatible with the normalisation of the large volume Yukawa couplings.
%%%%
\subsubsection{Further one-parameter models in this class}
Among the GLSMs associated to the $14$ one-parameter complete intersections in toric ambient spaces, there are two further models with hybrid phases of the same type. Their $I$-functions were computed in \cite{Erkinger:2020cjt}. Since the models are very similar, we only state the results.
\subsubsection*{Model K2}
The U(1) GLSM associated to this model has the following matter content.
\begin{equation}
  \begin{array}{c|ccc|c}
    &p_1,p_2&x_1,\ldots,x_4&x_5,x_6&\mathrm{FI}\\
    \hline
    U(1)&-4&1&2&\zeta\\
    U(1)_V&2-8q&2q&4q&
    \end{array}
\end{equation}
The large volume phase is the codimension two complete intersection $\mathbb{P}^5_{11111122}[4,4]$ of two quartics in weighted $\mathbb{P}^5$. Choosing $q=\frac{1}{4}$, the good hybrid phase at $\zeta\ll 0$ has $B=\mathbb{P}^1_{44}$ with $y^I=\{p_1,p_2\}$ and fibre coordinates $\phi^i=\{x_1,\ldots,x_6\}$ so that $X=\mathcal{O}(-1)^{\oplus 4}\oplus \mathcal{O}(-2)^{\oplus 2}$. The orbifold group is $\mathbb{Z}_4$ and the superpotential is $W=\langle p_1\rangle G_4^1(x_1,\ldots,x_6)+\langle p_2\rangle G_4^2(x_1,\ldots,x_6)$. Analysing the state space, we identify the sectors $\delta=2k=1,3$ to be the narrow sectors. The states in these sectors have exactly the same structure as those in (\ref{33states}). In particular, there is a state $|\Psi_1^0\rangle_{\frac{1}{2},-\frac{1}{2}}$ in the $\delta=1$-sector that defines the coordinate transformation between the $I$-function and the $J$-function. Computing the $J$-function and extracting the invariants is straightforward. As in the model K1 the selection rule (\ref{rule1}) picks the correlators (\ref{k1rule1}). The selection rule (\ref{rule2clader}) also holds for this model, with $N=4$ and imposes the constraint 
\begin{equation}
   -2-\beta=0\mod 4\qquad\leftrightarrow \qquad \beta=4n-2, \quad n=1,2,\ldots
\end{equation}
Setting $N=4$ in the pairing (\ref{pairing}), the structure of the $J$-function is easily derived and one gets the same form as in (\ref{k1jfun}). Using the $I$-function from \cite{Erkinger:2020cjt} and the coordinate transformation is obtained from $t=\frac{I_1^{H}}{I_1^{\bf 1}}$, the invariants can be read off from $\frac{I_3^{H}}{I_1^{\bf 1}}$. The result is
\begingroup
\renewcommand{\arraystretch}{1.3}
  \begin{equation}
  \begin{array}{c|c|c|c|c|c}
    4n-2&2&6&10&14&18\\
    \hline
    \langle\cdot\rangle_{4n-2}&\frac{1}{128}&\frac{1}{33554432}&\frac{237061}{237494511599616000}&\frac{3355}{55340232221128654848}&\frac{189816289475}{38692899125398935199712664354816}
    \end{array}
\end{equation}
 \endgroup
%%%%%
\subsubsection*{Model K3}
This hybrid model can be constructed as the $\zeta\ll0$-phase of the $U(1)$ GLSM with the following matter content.
\begin{equation}
  \begin{array}{c|cccc|c}
    &p_1,p_2&x_1,x_2&x_3,x_4&x_5,x_6&\mathrm{FI}\\
    \hline
    U(1)&-4&1&2&3&\zeta\\
    U(1)_V&2-12q&2q&4q&6q&
    \end{array}
\end{equation}
The geometric phase at $\zeta\gg0$ is a codimension two complete intersection $\mathbb{P}^5_{112233}[6,6]$ of two sextics in weighted $\mathbb{P}^5$. Setting $q=\frac{1}{6}$, the hybrid phase at $\zeta\ll0$ is a $G=\mathbb{Z}_6$ Landau-Ginzburg orbifold with base $\mathbb{P}^1_{66}$ parameterised by $y^I=\{p_1,p_2\}$ and fibre coordinates $\phi^i=\{x_1,\ldots,x_6\}$. This implies $X=\mathcal{O}(-1)^{\oplus 2}\oplus\mathcal{O}(-2)^{\oplus 2}\oplus\mathcal{O}(-3)^{\oplus 2}$ in the definition of the hybrid model. The superpotential is $W=\langle p_1\rangle G_6^1(x_1,\ldots,x_6)+\langle p_2\rangle G_6^2(x_1,\ldots,x_6)$. The narrow sectors are identified to be those with $\delta=2k=1,5$. Going through the steps to extract the invariants as for the models K1 and K2, we find:
\begingroup
\renewcommand{\arraystretch}{1.3}
 \begin{equation}
  \begin{array}{c|c|c|c|c}
    6n-2&4&10&16&22\\
    \hline
    \langle\cdot\rangle_{6n-2}&\frac{1}{55296}&\frac{1}{278628139008}&\frac{1251305}{368040959274957611728896}&\frac{1239223}{231812806445087701493115518976}%&\frac{8899960776291731}{820526770529616947509578703401952500252672000}
    \end{array}
 \end{equation}
 \endgroup
%%%%%%%%%%%%%%%%%%%%%%%%%%%%%%%%%%%%%%%%%%%%%%%%%%%%%%%%%%%%%%%%%%%%%%%%%
 \subsection{Model M1}
 \label{sec-m1}
Or next example is another model that is well-studied in the mathematics and physics literature, see for instance \cite{Caldararu:2007tc,Sharpe:2012ji,MR3590512} for accounts relevant in the present context. The model can be characterised by a $U(1)$ GLSM with matter content
\begin{equation}
  \begin{array}{c|cc|c}
    &p_1,\ldots,p_4&x_1,\ldots,x_8&\mathrm{FI}\\
    \hline
    U(1)&-2&1&\zeta\\
    U(1)_V&2-4q&2q&
    \end{array}
\end{equation}
The $\zeta\gg0$ phase is a complete intersection of four quadrics in $\mathbb{P}^7$. The vacuum in the $\zeta\ll0$ phase is a $\mathbb{P}^3_{2222}$ and the gauge symmetry is broken to $\mathbb{Z}_2$. Turning on fluctuations of the $x_i$, the resulting theory is massive unless the mass matrix $A_{ij}(p)=\sum_{k=1}^7A_{ij}^kp_k$ drops in rank. There is a non-trivial low energy theory as a branched double cover with branching locus $\mathrm{det}A_{ij}(p)=0$. The resulting phase of the GLSM has been interpreted in \cite{Caldararu:2007tc} as a non-commutative resolution of the singularity arising at the branching locus. Indeed, this model shares many properties of a ``typical'' geometric phase, some of which we will address below.

However, this model also fits into the framework of good hybrids, since the $\zeta\ll0$-phase can be viewed as a hybrid with base $B=\mathbb{P}^3_{2222}$ spanned by $y^I=\{p_1,\ldots,p_4\}$, above which a $G=\mathbb{Z}_2$ Landau-Ginzburg orbifold with superpotential $W=\sum_{i=1}^4\langle p_i\rangle G_2^i(x_1,\ldots,x_8)$ is fibered. Here, $G_2^i$ are quadrics in the fibre coordinates $\phi^i=\{x_1,\ldots,x_8\}$ and $X=\mathcal{O}(-1)^{\oplus 8}$. This is also how this model is characterised in the FJRW literature \cite{MR3590512}. In the following we proceed by treating the model as a good hybrid.  
%%%%%
\subsubsection{State space and narrow sector}
The narrow state space of this model has a somewhat richer structure compared to the previous examples. As before, we search for twisted sectors where all the fibre fields are heavy and the state space is defined in terms of the base. Since we only have four twisted sectors and all the fibre fields have $q_i=\frac{1}{2}$, the analysis is straightforward. The vacua have the following charges
\begingroup
\renewcommand{\arraystretch}{1.2}
  \begin{equation}
    \begin{array}{c|rrrr}
      k&0&1&2&3\\
      \hline
      q&-\frac{3}{2}&0&-\frac{3}{2}&0\\
      \hline
      \overline{q}&-\frac{3}{2}&-\frac{3}{2}&-\frac{3}{2}&-\frac{3}{2}
      \end{array}
  \end{equation}
  \endgroup
  Furthermore we compute the following values $\nu_i,\tilde{\nu}_i,\tau_i$ for the fibre fields:
  \begingroup
  \renewcommand{\arraystretch}{1.2}
  \begin{equation}
    \begin{array}{c|rrrr}
      k&0&1&2&3\\
      \hline
      \nu_i&0&\frac{1}{4}&\frac{1}{2}&\frac{3}{4}\\
      \hline
      \tilde{\nu}_i&0&-\frac{1}{4}&-\frac{1}{2}&-\frac{3}{4}\\
      \hline
      \tau_i&0&\frac{1}{2}&1&\frac{3}{2}
    \end{array}
  \end{equation}
  \endgroup
  The only sector with $k$ even and only heavy fibre fields is the one with $k=2$. Since we have a fibration over $\mathbb{P}^3$, we can turn on three $\overline{\eta}^I$ and three $\overline{\chi}_I$.
  This gives quite a few states on a local patch on $B$.
  \begingroup
  \renewcommand{\arraystretch}{1.3}
  \begin{equation}
    \begin{array}{llll}
    |\Psi_0^0\rangle_{-\frac{3}{3},-\frac{3}{2}} & |\Psi_1^0\rangle_{-\frac{3}{2},-\frac{1}{2}}^{\oplus 3} &|\Psi_2^0\rangle_{-\frac{3}{2},\frac{1}{2}}^{\oplus 3} &|\Psi_3^0\rangle_{-\frac{3}{2},\frac{3}{2}}\\
    |\Psi_0^1\rangle_{-\frac{1}{2},-\frac{3}{2}}^{\oplus 3} & |\Psi_1^1\rangle_{-\frac{1}{2},-\frac{1}{2}}^{\oplus 9} &|\Psi_2^1\rangle_{-\frac{1}{2},\frac{1}{2}}^{\oplus 9} &|\Psi_3^1\rangle_{-\frac{1}{2},\frac{3}{2}}^{\oplus 3}\\
    |\Psi_0^2\rangle_{\frac{1}{2},-\frac{3}{2}}^{\oplus 3} & |\Psi_1^2\rangle_{\frac{1}{2},-\frac{1}{2}}^{\oplus 9} &|\Psi_2^2\rangle_{\frac{1}{2},\frac{1}{2}}^{\oplus 9} &|\Psi_3^2\rangle_{\frac{1}{2},\frac{3}{2}}^{\oplus 3}\\
    |\Psi_0^3\rangle_{\frac{3}{2},-\frac{3}{2}} & |\Psi_1^3\rangle_{\frac{3}{2},-\frac{1}{2}}^{\oplus 3} &|\Psi_2^3\rangle_{\frac{3}{2},\frac{1}{2}}^{\oplus 3} &|\Psi_3^3\rangle_{\frac{3}{2},\frac{3}{2}}
    \end{array}
  \end{equation}
  \endgroup
The vacuum $|2\rangle$ is a section of $L_{|k\rangle}=\mathcal{O}(8)$ which is the anti-canonical bundle of $B$. The $\overline{\chi}_I$ and $\overline{\eta}^{\overline{I}}$ are sections of $T^*_B$ and $\overline{T}_B$ which is $\mathcal{O}(-2)$, we need three insertions of either $\overline{\chi}_I$ or $\overline{\eta}^{\overline{I}}$ on a local patch of $B$ to obtain an invariant state. As in the previous examples, this removes all states with $|q|\neq|\overline{q}|$. Since $B$ is a projective space, the three $\overline{\chi}_I$ and $\overline{\eta}^{\overline{I}}$, respectively, are equivalent in $\overline{Q}_0$ cohomology. The invariant states that contribute to the $Q_0$ cohomology are thus
\begin{equation}
  |\Psi_3^0\rangle_{-\frac{3}{2},\frac{3}{2}} \quad |\Psi_2^1\rangle_{-\frac{1}{2},\frac{1}{2}} \quad |\Psi_1^2\rangle_{\frac{1}{2},-\frac{1}{2}} \quad |\Psi_0^3\rangle_{\frac{3}{2},-\frac{3}{2}}.
\end{equation}
We can identify these states with elements of $H^*(\mathbb{P}^3)$ and we denote the basis of the associated $(a,c)$ state space by $e_{\delta,a}$ where $\delta=\frac{k}{2}=1$ and $a\in\{{\bf 1}, H,H^2,H^3\}$, with $H$ being the hyperplane class of $\mathbb{P}^3$. The pairing (\ref{pairing}) is then almost the geometric pairing, except for a normalisation factor $\frac{1}{2}$ coming from the $\mathbb{Z}_2$-orbifold, consistent with the gerby nature of this model.  
%%%%%%
\subsubsection{$J$-function and invariants}
We recall the $I$-function of this model \cite{MR3590512,Erkinger:2020cjt} that arises in the sector $\delta=1$, consistent with the analysis of the state space:
\begin{align}
\begin{split}
I^{\zeta \ll 0}_1(\mathsf{t},H) &= \frac{	\Gamma\left(1+ \frac{H}{2\pi i} \right)^{4}}{\Gamma\left(\frac{H}{2 \cdot 2 \pi i} +   \frac{1}{2} \right)^{8}} \sum_{a=0}^\infty
e^{\mathsf{t}(\frac{H}{2 \cdot 2 \pi i} +a)}
(-1)^{8a }
\frac{ \Gamma \left(a+ \frac{H}{2 \cdot 2 \pi i} +\frac{1}{2}\right)^{8}
}{
	\Gamma \left(1 +2a +\frac{H}{2\pi i} \right)^{4}}.
\end{split}
\label{eqn:M1IFunc}
\end{align}
Expanding this expression in $H$, we get a schematic expansion consistent with the cohomology $H^*(\mathbb{P}^3)$:
\begin{equation}
  I_1(\mathsf{t})=I_1^{\bf{1}}e_{1,{\bf 1}}+I_1^{H}e_{1,H}+I_1^{H^2}e_{1,H^2}+I_1^{H^3}e_{1,H^3}.
\end{equation}
This is what one would expect in a geometric phase, the only difference being that the $I$-function arises in a twisted sector. Matching the basis elements with the corresponding states in the hybrid conformal field theory, we can identify the coordinate transformation to obtain the $J$-function:
\begin{equation}
  t(\mathsf{t})=\frac{I_1^H(u(\mathsf{t}))}{I_1^{\bf}(u(\mathsf{t}))},
\end{equation}
where we set $u(\mathsf{t})=e^{\mathsf{t}}$. By the selection rule (\ref{rule1}), the correlators appearing in the $J$-function have the same form as (\ref{k1rule1}) for the K-type models above. The selection rule (\ref{rule2clader}) is also applicable for this model and now reads
\begin{equation}
  -2-\beta=0\mod 2\qquad\leftrightarrow \qquad \beta=2n-2, \quad n=1,2,\ldots
\end{equation}
Note that, in contrast to the previous examples, this includes the case $\beta=0$. General properties of geometric models imply that the only genus-$0$ correlation functions with $\beta=0$ are the three-point correlators $\langle \phi_{i_1}\phi_{i_2}\phi_{i_3}\rangle_{0,3,0}$. In the present example, the only three-point function that can appear in the $J$-function is $\langle e_{1,H} e_{1,H} e_{1,H}\rangle_{0,3,0}$ and we get a slightly modified structure:
\begin{align}
  \label{m1jfun}
  J=&e_{1,{\bf 1}}+te_{1,H}+2e_{1,H^2}\left[\frac{t^2}{2}\langle e_{1,H}e_{1,H}e_{1,H}\rangle_{0,3,0}+\sum_{n\geq 2}e^{(2n-2)t}(2n-2)\langle\cdot\rangle_{0,0,2n-2}\right]\nonumber\\
  &+2e_{1,H^3}\left[\frac{t^3}{6}\langle e_{1,H}e_{1,H}e_{1,H} \rangle_{0,3,0}+\sum_{n\geq 2}e^{(2n-2)t}[(2n-2)t-2]\langle\cdot\rangle_{0,0,2n-2}\right].
\end{align}
The coordinate $t$ is given by
\begin{equation}
  t=\frac{I_1^H(u)}{I_1^{\bf 1}(u)}=\log u+\frac{u^2}{2048}+\frac{157 u^4}{268435456}+\frac{917
   u^6}{824633720832}+\frac{5944509 u^8}{2305843009213693952}+O\left(u^{9}\right)
\end{equation}
From this we obtain with $q=e^t$
\begin{equation}
  u(q)=q-\frac{q^3}{2048}+\frac{3 q^5}{268435456}-\frac{35 q^7}{549755813888}-\frac{167853q^9}{2305843009213693952}+O(q^{11})%-\frac{511825q^{11}}{4722366482869645213696}
  \end{equation}
Inserting into $\frac{I_1^H}{I_1^{\bf 1}}$ one finds
\begin{equation}
 \frac{I_1^H(u)}{I_1^{\bf 1}(u)} =\frac{\log ^2q}{2}+\frac{q^2}{8192}+\frac{153 q^4}{2147483648}+\frac{343
   q^6}{4947802324992}+\frac{3213145 q^8}{36893488147419103232}+O\left(q^{9}\right)
\end{equation}
Comparing with (\ref{m1jfun}), the result for the three-point function is 
\begin{equation}
  \langle e_{1,H}e_{1,H}e_{1,H}\rangle_{0,3,0}=\frac{1}{2}.
\end{equation}
In the case of Calabi-Yau threefolds, this correlator is related to the triple intersection number, so its appearance here is consistent with the geometric interpretation of this hybrid model. Assuming that the normalisation for the pairing (\ref{pairing}) is correct, the fact that this number is not integer seems consistent with the $\mathbb{Z}_2$-gerbe structure of this model. 
For the first few invariants we get
\begingroup
\renewcommand{\arraystretch}{1.3}
\begin{equation}
  \begin{array}{c|c|c|c|c|c}
    2n-2&2&4&6&8&10\\
    \hline
    \langle\cdot\rangle_{2n-2}&\frac{1}{32768}&\frac{153}{17179869184}&\frac{343}{59373627899904}&\frac{3213145}{590295810358705651712}&\frac{29876461}{4722366482869645213696000}
    \end{array}
\end{equation}
\endgroup
%%%%%
\subsubsection{Relation to the integer invariants computed in \cite{Sharpe:2012ji}}
While from a mathematical and GLSM perspective this model is a hybrid model, it has been shown in \cite{Caldararu:2007tc} that it can be given a geometric interpretation. In \cite{Sharpe:2012ji}, the prescription of \cite{Jockers:2012dk} to extract instanton numbers from the GLSM sphere partition function for geometric phases has been used to compute these integer invariants. We expect that our results for the FJRW invariants should be related to these numbers in the same way that Gromov-Witten invariants are related to the integer instanton numbers. This is indeed the case as we shall now demonstrate.

In \cite{Sharpe:2012ji}, the following integer invariants have been computed using the sphere partition function:
\begin{equation}
  \begin{array}{c|ccccc}
    i&1&2&3&4&5\\
    \hline
    N_i&64&1216&52032&3212992&244747968
    \end{array}
  \end{equation}
We have also confirmed these numbers by a mirror symmetry calculation as discussed in Section~\ref{sec-mir33}. The FJRW invariants and instanton numbers should be related by undoing multicoverings. Writing $\langle\cdot\rangle_{2n-2}=n_{2n-2}$, we obtain the following relations: 
\begin{align}
  {2^{16}}2^51^3n_2=&N_1\nonumber\\
  {2^{32}}2^52^3n_4=&N_1+8N_2\nonumber\\
  {2^{48}}2^53^3n_6=&N_1+27N_3\nonumber\\
  {2^{64}}2^54^3n_8=&N_1+8N_2+64N_4\nonumber\\
  {2^{80}}2^55^3n_{10}=&N_1+125N_5\nonumber\\
  \ldots=&\ldots\nonumber\\
  {2^{16k}}2^5k^3n_{2k}=&f_k(N),
\end{align}
where the expansion of the instanton numbers (see e.g.~\cite{Cox:2000vi}) $\sum_{a=0}^{\infty}\frac{a^3 N_aq^a}{1-q^a}=\sum_{k=0}^{\infty}f_k(N)q^k$ determines the $f_k(N)$. 
%%%%%%%%%%%%%%%%%%%%%%%%%%%%%%%%%%%%%%%%%%%%%%%%%%%%%%%%%%%%%%%%%%%%%%%%%%%%%%
\section{Two-parameter examples}
\label{sec-2par}
In this section we compute FJRW invariants for hybrid phases of two well-studied two-parameter GLSMs. In contrast to the one-parameter models, examples of this type have, to our knowledge, not been discussed in the mathematics literature. 
\subsection{Example 1}
We have discussed this model in \cite{Erkinger:2020cjt}, so we only recall the necessary results. The GLSM associated to this model has $\mathsf{G}=U(1)^2$ and the following matter content:
\begin{equation}
  \begin{array}{c|rr|rrrrr|c}
    &p&x_6&x_3&x_4&x_5&x_1&x_2&\mathrm{FI}\\
    \hline
    U(1)_1&-4&1&1&1&1&0&0&\zeta_1\\
    U(1)_2&0&-2&0&0&0&1&1&\zeta_2\\
    \hline
    U(1)_V&2-8q_1&2q_1-4q_2&2q_1&2q_1&2q_1&2q_2&2q_2,
    \end{array}
\end{equation}
where $0\leq q_1\leq\frac{1}{4}$ and $0\leq q_2\leq \frac{1}{8}$. The GLSM superpotential is $W=pG_{(4,0)}(x_1,\ldots,x_6)$. The subscript gives the degree of $G$ with respect to the weights given by the $U(1)$-charges. The large volume phase at $\zeta_1\gg0,\zeta_2\gg0$ is a two-parameter Calabi-Yau hypersurface in the ambient toric variety defined by the $U(1)^2$-charges of the $x_i$. The Calabi-Yau is a K3-fibration over $\mathbb{P}^1$. 

The hybrid phase of this model has also been one of the main examples in \cite{Bertolini:2013xga,Bertolini:2018now}. In the moduli space of the GLSM it is located in the limiting region $\zeta_1\ll0,\zeta_2\gg0$. In this phase, $p$ obtains a VEV and the base $B$ of the hybrid model is a $\mathbb{P}^1$ parameterised by $y^I=\{x_1,x_2\}$. In this example, the base is a true $\mathbb{P}^1$ rather than a ``gerby'' $\mathbb{P}^1$. The Landau-Ginzburg orbifold fibred over this base manifold has $G=\mathbb{Z}_4$ with fibre coordinates $\phi^i=\{x_3,\ldots,x_6\}$. The bundle $X$ in the definition of the hybrid model is $X=\mathcal{O}(-2)\oplus\mathcal{O}^{\oplus 3}$. To match with the vector R-charges in the low energy theory, we have to choose $q_1=\frac{1}{4},q_2=0$. 
%%%%%
\subsubsection{State space and narrow sector}
To find the narrow state space, we proceed exactly as for the one-parameter models. All the fibre fields have R-charge $q_i=\frac{1}{4}$. The vacua $|k\rangle$ have the following left and right R-charges:
\begingroup
\renewcommand{\arraystretch}{1.2}
\begin{equation}
  \begin{array}{c|rrrrrrrr}
    k&0&1&2&3&4&5&6&7\\
    \hline
    q&-\frac{3}{2}&0&\frac{1}{2}&-1&-\frac{1}{2}&1&-\frac{3}{2}&0 \\
    \hline
    \overline{q}&-\frac{3}{2}&-\frac{3}{2}&-\frac{3}{2}&-\frac{1}{2}&-\frac{1}{2}&-\frac{1}{2}&\frac{1}{2}&\frac{1}{2}
    \end{array}
\end{equation}
\endgroup
Furthermore one finds:
\begingroup
\renewcommand{\arraystretch}{1.3}
\begin{equation}
   \begin{array}{c|rrrrrrrr}
    k&0&1&2&3&4&5&6&7\\
    \hline
    \nu_i&0&\frac{1}{8}&\frac{1}{4}&\frac{3}{8}&\frac{1}{2}&\frac{5}{8}&\frac{3}{4}&\frac{7}{8}\\
    \hline
    \overline{\nu}_i& 0&-\frac{3}{8}&-\frac{3}{4}&-\frac{1}{8}&-\frac{1}{2}&-\frac{7}{8}&-\frac{1}{4}&-\frac{5}{8}\\
    \hline
    \tau_i&0&\frac{1}{2}&1&\frac{1}{2}&1&\frac{3}{2}&1&\frac{3}{2}
    \end{array}
  \end{equation}
\endgroup
We identify the $k\in \mathbb{Z}_8$-sectors with $k=2\delta=2,4,6$ as the narrow sectors. In these sectors, all the fibre fields are heavy and the corresponding vacua are invariant under the action of the orbifold. They transform as sections of $L_{|k\rangle}=\mathcal{O}(2)\otimes\mathcal{O}^{\otimes 3}=\mathcal{O}(2)$. This is the anti-canonical bundle of $B$. To compensate for this, we need to turn on $\overline{\chi}_I$ or $\overline{\eta}^{\overline{I}}$ to obtain an invariant state. This leads to six states with $q=-\overline{q}$ that are related, via spectral flow, to the narrow $(a,c)$-ring elements:
\begin{align}
  k&=2:&|\Psi_{0}^1\rangle_{\frac{3}{2},-\frac{3}{2}}&&|\Psi_{1}^0\rangle_{\frac{1}{2},-\frac{1}{2}}\nonumber\\
  k&=4:& |\Psi_{0}^1\rangle_{\frac{1}{2},-\frac{1}{2}}&&|\Psi_{1}^0\rangle_{-\frac{1}{2},\frac{1}{2}}\nonumber\\
   k&=6:&|\Psi_{0}^1\rangle_{-\frac{1}{2},\frac{1}{2}}&&|\Psi_{1}^0\rangle_{-\frac{3}{2},\frac{3}{2}}
  \end{align}
As expected, these states combine into three copies of $H^*(\mathbb{P}^1)$ and we denote the corresponding basis elements of the state space by $e_{\delta,a}$ with $\delta\in\{1,2,3\}$ and $a\in\{1,H\}$. The charges indicate that the elements $e_{1,H}$ and $e_{2,{\bf 1}}$ define the coordinate transformation between the $I$-function and the $J$-function.

Making use of (\ref{pairing}), we have the following non-trivial pairings between the states:
\begin{equation}
  \label{deg8pairing}
  \langle e_{1,{\bf 1}},e_{3,H}\rangle=\langle e_{1,H},e_{3,{\bf 1}}\rangle=\langle e_{2,{\bf 1}},e_{2,H}\rangle=\frac{1}{4}.
  \end{equation}
%%%%%
\subsubsection{$J$-function and invariants}
Based on the GLSM sphere partition function, a proposal for the  $I$-function for this model was made in \cite{Erkinger:2020cjt}:
\begin{equation}
  I_{\delta}(\mathsf{t}_1,\mathsf{t}_2,H)=\frac{\Gamma\left(1+\frac{H}{2\pi i}\right)^2}{\Gamma\left(\frac{\delta}{4}+\frac{H}{\pi i}\right)\Gamma\left(\frac{\delta}{4}\right)^3}e^{-\mathsf{t}_2\frac{H}{2\pi i}}\sum_{a,n\geq 0}\frac{\Gamma\left(a+\frac{\delta}{4}+2n+2\frac{H}{2\pi i} \right)\Gamma\left(a+\frac{\delta}{4} \right)^3}{\Gamma\left(4a+\delta\right)\Gamma\left(1+n+\frac{H}{2\pi i} \right)^2}e^{\frac{\mathsf{t}_1}{4}(4a+\delta-1)}e^{-\mathsf{t}_2n}.
\end{equation}
Expanding further to linear order in $H$, we can write the $I$-function as
\begin{equation}
  I(\mathsf{t}_1,\mathsf{t}_2)=\sum_{\delta=1}^3\sum_{a\in\{{\bf 1},H\}}I_{\delta}^ae_{\delta,a},
  \end{equation}
with the $e_{\delta,a}$ defined via the narrow state space as above. From this we can read off the definition of the flat coordinates:
\begin{equation}
  \label{deg8tdef}
  t_1=\frac{I_{1,H}(u_1(\mathsf{t}_1,\mathsf{t}_2),u_2(\mathsf{t}_1,\mathsf{t}_2))}{I_{1,{\bf 1}}(u_1(\mathsf{t}_1,\mathsf{t}_2),u_2(\mathsf{t}_1,\mathsf{t}_2))}, \qquad t_2=\frac{I_{2,{\bf 1}}(u_1(\mathsf{t}_1,\mathsf{t}_2),u_2(\mathsf{t}_1,\mathsf{t}_2))}{I_{1,{\bf 1}}(u_1(\mathsf{t}_1,\mathsf{t}_2),u_2(\mathsf{t}_1,\mathsf{t}_2))},
  \end{equation}
where we identify
\begin{equation}
  u_1(\mathsf{t}_1,\mathsf{t}_2)=e^{-\mathsf{t}_2}, \qquad u_2(\mathsf{t}_1,\mathsf{t}_2)=e^{\frac{\mathsf{t}_1}{4}}. 
\end{equation}
Using the selection rule (\ref{rule1}), the expansion of the $J$-function contains the following correlators:
\begin{equation}
  \label{deg8corr}
  \langle \tau_1(e_{1,{\bf 1}})(e_{1,H})^{k_1}(e_{2,{\bf 1}})^{k_2}\rangle_{0,k_1+k_2+1,\beta}, \quad\langle(e_{1,H})^{k_1+1}(e_{2,{\bf 1}})^{k_2}\rangle_{0,k_1+k_2+1,\beta},\quad \langle(e_{1,H})^{k_1}(e_{2,{\bf 1}})^{k_2+1}\rangle_{0,k_1+k_2+1,\beta}.
\end{equation}
This, together with the pairing (\ref{deg8pairing}), allows us to write the $J$-function as
\begin{align}
  J=&e_{1,{\bf 1}}+t_1e_{1,H}+t_2e_{2,{\bf 1}}\nonumber\\
  &+4e_{3,H}\sum_{k_1,k_2,\beta\geq0}\frac{t_1^{k_1}}{k_1!}\frac{t_2^{k_2}}{k_2!}\langle \tau_1(e_{1,{\bf 1}})(e_{1,H})^{k_1}(e_{2,{\bf 1}})^{k_2}\rangle_{0,k_1+k_2+1,\beta}\nonumber\\
  &+4e_{3,{\bf 1}}\sum_{k_1,k_2,\beta\geq0}\frac{t_1^{k_1}}{k_1!}\frac{t_2^{k_2}}{k_2!}\langle (e_{1,H})^{k_1+1}(e_{2,{\bf 1}})^{k_2}\rangle_{0,k_1+k_2+1,\beta}\nonumber\\
  &+4e_{2,H}\sum_{k_1,k_2,\beta\geq0}\frac{t_1^{k_1}}{k_1!}\frac{t_2^{k_2}}{k_2!}\langle (e_{1,H})^{k_1}(e_{2,{\bf 1}})^{k_2+1}\rangle_{0,k_1+k_2+1,\beta}.
\end{align}
Next, we consider the selection rule (\ref{rule2}). In contrast to the previous examples, the base coordinates do not transform under the orbifold symmetry, and hence $q_B(\beta)=0$. The selection rule thus reduces to
\begin{equation}
  -2+m-\sum_{i=1}^m\delta_{i}=0\mod 4,
\end{equation}
with $m$ the number of insertions and $\delta_i$ the label of the twisted sector of the $i$-th insertion. Applying this to all three types of correlators, we obtain the following conditions: 
\begin{align}
  \label{deg8rule2}
  \text{(a)}:&\quad k_2=2\mod 4 \nonumber \\
  \text{(b)}:&\quad k_2=1\mod 4 
\end{align}
where (a) applies to the first two types of correlators on (\ref{deg8corr}), i.e.~those containing $k_2$ insertions of $e_{2,{\bf 1}}$, and (b) applies to the one with $k_2+1$ insertions of $e_{2,{\bf 1}}$. 
Furthermore, we can use the dilaton equation to write
\begin{equation}
  \langle \tau_1(e_{1,{\bf 1}})(e_{1,H})^{k_1}(e_{2,{\bf 1}})^{k_2}\rangle_{0,k_1+k_2+1,\beta}=(-2+k_1+k_2)\langle(e_{1,H})^{k_1}(e_{2,{\bf 1}})^{k_2}\rangle_{0,k_1+k_2,\beta}.
\end{equation}
Since $e_{1,H}\in H^2(B)$, we can apply the divisor equation to this insertion if $\beta\neq 0$ to obtain
\begin{equation}
  \langle(e_{1,H})^{k_1}(e_{2,{\bf 1}})^{k_2}\rangle_{0,k_1+k_2,\beta}=\left(\int_{\beta}e_{1,H}\right)^{k_1}\langle(e_{2,{\bf 1}})^{k_2}\rangle_{0,k_2,\beta}=\beta^{k_1}\langle(e_{2,{\bf 1}})^{k_2}\rangle_{0,k_2,\beta}.
\end{equation}
where $k_2$ is subject to the selection rule (\ref{deg8rule2}). For $\beta=0$ we can have the following correlators
\begin{equation}
  \langle e_{1,H}(e_{2,{\bf 1}})^2\rangle_{0,3,0}, \qquad \langle (e_{2,{\bf 1}})^{k_2}\rangle_{0,k_2,0}.
\end{equation}
Using this information, the $J$-function can be written as
\begin{align}
  J=&e_{1,{\bf 1}}+t_1e_{1,H}+t_2e_{2,{\bf 1}}\nonumber\\
  &+4e_{2,H}\left[t_1t_2\langle e_{1,H}(e_{2,{\bf 1}})^2\rangle_{0,3,0}+\sum_{k_2,\beta\geq 0}e^{\beta t_1}\frac{t_2^{k_2}}{k_2!}\langle (e_{2,{\bf 1}})^{k_2+1}\rangle_{0,k_2+1,\beta} \right]\nonumber\\
  &+4e_{3,{\bf 1}}\left[\frac{t_2^2}{2}\langle e_{1,H}(e_{2,{\bf 1}})^2\rangle_{0,3,0}+\sum_{k_2\geq0,\beta>0}\beta e^{\beta t_1}\frac{t_2^{k_2}}{k_2!}\langle (e_{2,{\bf 1}})^{k_2}\rangle_{0,k_2,\beta}\right]\nonumber\\
  &+4e_{3,H}\left[t_1\frac{t_2^2}{2}\langle e_{1,H}(e_{2,{\bf 1}})^2\rangle_{0,3,0}+\sum_{k_2,\beta\geq0}e^{\beta t_1}\frac{t_2^{k_2}}{k_2!}(-2+k_2+\beta t_1)\langle (e_{2,{\bf 1}})^{k_2}\rangle_{0,k_2,\beta}\right].
\end{align}
All the invariants can be read off from the $e_{2,H}$-component of the $J$-function. The coordinate transformation (\ref{deg8tdef}) gives
\begin{align}
  \label{deg8mir}
  t_1=&\log u_1+\frac{19}{8} u_1+\frac{3981u_1^2}{1024}+\frac{112661 u_1^3}{12288}+\frac{\left(633749415u_1^4+32768 u_2^4\right)}{25165824}\nonumber\\
  &+\frac{\left(18935316729 u_1^5+778240u_2^4 u_1\right)}{251658240}+\ldots\nonumber\\
  t_2=&u_2+\frac{7}{16} u_1 u_2+\frac{955 u_1^2u_2}{1024}+\frac{41845u_1^3 u_2}{16384}+\left(\frac{11u_2^5}{30720}+\frac{32652715 u_1^4 u_2}{4194304}\right)+\ldots
\end{align}
Defining $q_1=e^{t_1}$ and $q_2=t_2$, this can be inverted to give
\begin{align}
  u_1&=q_1-\frac{19}{8}q_1^2+\frac{4683q_1^3}{1024}-\frac{32601q_1^4}{4096}+\left(\frac{109687655q_1^5}{8388608}-\frac{1}{768} q_1 q_2^4\right)+\ldots \nonumber\\
  u_2&=q_2-\frac{7}{16} q_1 q_2+\frac{305 q_1^2q_2}{1024}-\frac{309q_1^3 q_2}{1024}+\left(\frac{1351233 q_1^4 q_2}{4194304}-\frac{11 q_2^5}{30720}\right)+\ldots
\end{align}
The $e_{2,H}$-term of the $J$-function then becomes
\begin{equation}
  \frac{I_2^H}{I_1^{\bf 1}}=q_2 \log q_1+\frac{1}{8} q_1 q_2+\frac{19 q_1^2 q_2}{1024}+\frac{97 q_1^3 q_2}{24576}+\frac{\left(32768q_2^5+40975 q_1^4 q_2\right)}{41943040}+\ldots
\end{equation}
From the first term we obtain
\begin{equation}
  \langle e_{1,H}(e_{2,{\bf 1}})^2\rangle_{0,3,0}=\frac{1}{4}.
\end{equation}
The other invariants are of the form, $\langle (e_{2,{\bf 1}})^n\rangle_{0,n,\beta}$ with $n=2\mod 4$. From this, we compute
\begingroup
\renewcommand{\arraystretch}{1.3}
\begin{equation}
  \begin{array}{c|ccccc}
    \beta\backslash n&2&6&10&14&18\\
    \hline
    0&-&\frac{3}{128}&\frac{9}{256}&\frac{451521}{1048576}&\frac{11035197}{524288}\\
    1&\frac{1}{32}&\frac{9}{512}&\frac{1431}{8192}&\frac{30899253}{4194304}&\frac{15499432863}{16777216}\\
    2&\frac{19}{4096}&\frac{333}{32768}&\frac{2133}{8192}&\frac{6325575501}{268435456}&\frac{23182643961}{4194304}\\
    3&\frac{97}{98304}&\frac{2751}{524288}&\frac{2196549}{8388608}&\frac{178324775307}{4294967296}&\frac{270849007998117}{17179869184}
    \end{array}
\end{equation}
\endgroup
The other components of the $I$-function yield redundant information. 
%%%%%
\subsubsection{Mirror symmetry check}
Mirror symmetry for the geometric phase of this model has been discussed at length in \cite{Candelas:1993dm,Hosono:1993qy}, or in \cite{Cox:2000vi}. We can proceed as for the one-parameter models and check if the Yukawa couplings computed on the mirror are consistent with the form (\ref{jprepot}).

First we recall the Picard-Fuchs operators associated to the large complex structure limit of the mirror Calabi-Yau:
\begin{align}
  \mathcal{L}_1&=\theta_1^2(\theta_1-2\theta_2)-4z_1(4\theta_1+3)(4\theta_1+2)(4\theta_1+1),\nonumber\\
  \mathcal{L}_2&=\theta_2^2-z_2(2\theta_2-\theta_1+1)(2\theta_2-\theta_1),
\end{align}
where $\theta_i=z_i\frac{\partial}{\partial z_i}$ and we identify $z_1=e^{-\mathsf{t}_1}$ and $z_{2}=e^{-\mathsf{t}_2}$. The modulus $z_1$ is associated to the K3 fibre and $z_2$ is associated to the base. Choosing divisor classes $H$ for the fibre and $L$ for the base, the triple intersection numbers are
\begin{equation}
  H^3=8, \qquad H^2L=4, \qquad HL^2=0,\qquad L^3=0.
\end{equation}
The discriminant is
\begin{equation}
  \Delta=(1-256z_1)^2-512^2z_1^2z_2.
  \end{equation}
The Yukawa couplings, normalised in agreement with the triple intersection numbers, are:
\begin{align}
  Y_{z_1z_1z_1}&=\frac{8}{\Delta}, && Y_{z_1z_1z_2}=\frac{4(1-256z_1)}{\Delta}\nonumber\\
  Y_{z_1z_2z_2}&=\frac{8(1-512z_1)z_2}{(-1+4z_2)\Delta}, && Y_{z_2z_2z_2}=\frac{4z_2(1+4z_2-256z_1(1+12z_2))}{(1-4z_2)^4\Delta}.
\end{align}
To transform these couplings into the hybrid phase, we have to identify $z_1=u_2^{-4}$ and $z_2=u_1$ to match with our conventions. Note in particular that
\begin{equation}
  u_1\frac{\partial}{\partial u_1}=z_2\frac{\partial}{\partial z_2}, \qquad \frac{\partial}{\partial u_2}=(-4u_2^{-1})z_1\frac{\partial}{\partial z_1}. 
\end{equation}
We further rescale the holomorphic threeform $\Omega(z_1,z_2)\rightarrow u_2\Omega(u_1,u_2)$ to account for the shift in $R$-charge between the low-energy descriptions of the geometric and hybrid phases. 
This gives, for example,
\begin{equation}
  Y_{u_1u_2u_2}=u_2^{-2}(-4u_2^{-1})^2Y_{z_1z_1z_2}(z_1(u_1,u_2),z_2(u_1,u_2))=\frac{64(-256+u_2^4)}{(-265+u_2^4)^2-512^2u_1},
\end{equation}
and similarly for the other three Yukawa couplings.

Next, we apply the coordinate transformation (\ref{deg8mir}) which we reinterpret as the mirror map. Since this is a bit tricky we give a few more details. We write
\begin{equation}
  q_1\frac{\partial}{\partial q_1}=a_1u_1\frac{\partial}{\partial u_1}+a_2\frac{\partial}{\partial u_2}, \qquad \frac{\partial}{\partial q_2}=b_1u_1\frac{\partial}{\partial u_1}+b_2\frac{\partial}{\partial u_2},
\end{equation}
with
\begin{equation}
  a_1=\frac{q_1}{u_1}\frac{\partial u_1}{\partial q_1}, \quad a_2=q_1\frac{\partial u_2}{\partial q_1}, \quad b_1=\frac{1}{u_1}\frac{\partial u_1}{\partial q_2}, \quad b_2=\frac{\partial u_2}{\partial q_2}.
\end{equation}
With that, we get for example 
\begin{align}
  C_{t_1t_1t_2}&=\frac{1}{\varpi_0^2}\left(a_1^2b_1Y_{u_1u_1u_1}+(2a_1a_2b_2+a_1^2b_2)Y_{u_1u_1u_2}+(2a_1a_2b_2+a_2^2b_1)Y_{u_1u_2u_2}+a_2^2b_2Y_{u_2u_2u_2}\right)\nonumber\\
  &=\frac{q_1 q_2}{32}+\frac{19 q_1^2 q_2}{1024}+\frac{291 q_1^3 q_2}{32768}+\frac{8195 q_1^4 q_2}{2097152}+\frac{3 \left(5888200 q_1^5 q_2+524288 q_1 q_2^5\right)}{10737418240}+\ldots,
\end{align}
where we identify $\varpi_0=I_1^{\bf 1}$. We get similar expressions for the other Yukawa couplings.

Finally, we can compare with (\ref{jprepot}). We can consider for instance the $e_{2,H}$-coefficient of the $J$-function that we have used to extract the invariants. In accordance with expectations we find
\begin{equation}
  q_1\frac{\partial}{q_1}\left(q_1\frac{\partial}{\partial q_1}\left(\frac{1}{4}\frac{I_2^H}{I_1^{\bf 1}}\right)\right)=C_{t_1t_1t_2}, \quad  \frac{\partial}{q_2}\left(q_1\frac{\partial}{\partial q_1}\left(\frac{1}{4}\frac{I_2^H}{I_1^{\bf 1}}\right)\right)=C_{t_1t_2t_2}, \quad \frac{\partial^2}{\partial q_2^2}\left(\frac{1}{4}\frac{I_2^H}{I_1^{\bf 1}}\right)=C_{t_2t_2t_2}.
  \end{equation}
%%%%%
\subsection{Example 2}
\label{sec-2parex2}
We discuss one further model which is a Landau-Ginzburg orbifold fibration over $\mathbb{P}^2$ and thus has slightly different properties compared to the previous examples.

We consider the following $\mathsf{G}=U(1)^2$ GLSM:
\begin{equation}
  \begin{array}{c|rr|rrrrr|c}
    &p&x_6&x_4&x_5&x_1&x_2&x_3&\mathrm{FI}\\
    \hline
    U(1)_1&-6&1&2&3&0&0&0&\zeta_1\\
    U(1)_2&0&-3&0&0&1&1&1&\zeta_2\\
    \hline
    U(1)_V&2-12q_1&2q_1-6q_2&4q_1&6q_1&2q_2&2q_2&2q_2
    \end{array}
\end{equation}
where $0\leq q_1\leq\frac{1}{6}$ and $0\leq q_2\leq \frac{1}{18}$ and $W=pG_{(6,0)}(x_1,\ldots,x_6)$. The geometric phase $\zeta_1\gg0,\zeta_2\gg0$ is a well-studied elliptically fibered Calabi-Yau threefold given by $G_{(6,0)}=0$ in the toric variety defined by gauge charges of the $x$-fields. We focus on the hybrid phase at $\zeta_1\ll0,\zeta_2\gg0$ and $p$ gets a VEV which breaks to gauge symmetry to a $\mathbb{Z}_6$. This phase is a true hybrid model with $B=\mathbb{P}^2$ parameterised by $y^I=\{x_1,x_2,x_3\}$. The fibre coordinates are $\phi^i=\{x_6,x_4,x_5\}$ with charges $q^i=\{\frac{1}{6},\frac{1}{3},\frac{1}{2}\}$, respectively, and the bundle defining the hybrid model is $X=\mathcal{O}(-3)\oplus\mathcal{O}^{\oplus 2}$. This corresponds to choosing $q_1=\frac{1}{6},q_2=0$ in the GLSM. 
%%%%
\subsubsection{State space and narrow sector}
Performing the familiar analysis of the state space, we identify the $\mathbb{Z}_{12}$-twisted sectors with $k=2\delta=2,10$ as narrow sectors. The vacua in these sectors are sections of $L_{|k\rangle}=\mathcal{O}(3)$. To construct invariant states on a local patch of $B$, we thus need to insert two of the fermions $\overline{\chi}_I$ or $\overline{\eta}^I$. Taking the $\overline{Q}_0$-cohomology, we identify the following representatives for the elements of the $(a,c)$-ring:
    \begin{align}
      k=2:&\quad |\Psi_0^2\rangle_{\frac{3}{2},-\frac{3}{2}}, \quad |\Psi_1^1\rangle_{\frac{1}{2},-\frac{1}{2}}, \quad |\Psi_2^0\rangle_{-\frac{1}{2},\frac{1}{2}} \nonumber\\
      k=10:&\quad|\Psi_0^2\rangle_{\frac{1}{2},-\frac{1}{2}}, \quad |\Psi_1^1\rangle_{-\frac{1}{2},\frac{1}{2}}, \quad |\Psi_2^0\rangle_{-\frac{3}{2},\frac{3}{2}}
    \end{align}
As it should be, the contribution of each sector gives a copy of $H^*(\mathbb{P}^2)$ and we denote the corresponding basis elements of the $(a,c)$-ring by $e_{\delta,a}$ with $\delta\in\{1,5\}$ and $a\in\{{\bf 1},H,H^2\}$. 
    %%%%%%%
\subsubsection{$J$-function and invariants}
The $I$-function for this model was not stated explicitly in \cite{Erkinger:2020cjt} but a straightforward computation yields
\begin{align}
  I_{\delta}(\mathsf{t}_1,\mathsf{t}_2,H)=&\frac{\Gamma\left(1+\frac{H}{2\pi i}\right)^3}{\Gamma\left(\left\langle\frac{\delta}{6}\right\rangle+\frac{3H}{2\pi i}\right)\Gamma\left(\left\langle\frac{\delta}{3}\right\rangle\right)\Gamma\left(\left\langle\frac{\delta}{2}\right\rangle\right)}e^{-\mathsf{t}_2\frac{H}{2\pi i}}\nonumber\\
  &\cdot\sum_{a,n\geq 0}\frac{\Gamma\left(a+\frac{\delta}{6}+3n+\frac{3H}{2\pi i}\right)\Gamma\left(2a+\frac{\delta}{3}\right)\Gamma\left(3a+\frac{\delta}{2}\right)}{\Gamma(6a+\delta)\Gamma\left(1+n+\frac{H}{2\pi i}\right)^3}(-1)^ne^{\frac{\mathsf{t}_1}{6}(6a+\delta-1)}e^{-\mathsf{t}_2n}.
\end{align}
From this we can obtain an expansion of the form
\begin{equation}
  I(\mathsf{t}_1,\mathsf{t}_2)\sum_{\delta\in \{1,5\}}\sum_{a\in\{1,H,H^2\}}I_{\delta}^ae_{\delta,a}.
\end{equation}
The coordinates of the $J$-function are
\begin{equation}
  t_1=\frac{I_{1,H}(u_1(\mathsf{t}_1,\mathsf{t}_2),u_2(\mathsf{t}_1,\mathsf{t}_2))}{I_{1,{\bf 1}}(u_1(\mathsf{t}_1,\mathsf{t}_2),u_2(\mathsf{t}_1,\mathsf{t}_2))}, \qquad t_2=\frac{I_{5,{\bf 1}}(u_1(\mathsf{t}_1,\mathsf{t}_2),u_2(\mathsf{t}_1,\mathsf{t}_2))}{I_{1,{\bf 1}}(u_1(\mathsf{t}_1,\mathsf{t}_2),u_2(\mathsf{t}_1,\mathsf{t}_2))},
  \end{equation}
where we identify\footnote{The minus sign in the definition of $u_1$ accounts for the $\theta$-angle shift between the GLSM and the low-energy theory\cite{Herbst:2008jq}.} 
\begin{equation}
  u_1(\mathsf{t}_1,\mathsf{t}_2)=-e^{-\mathsf{t}_2}, \qquad u_2(\mathsf{t}_1,\mathsf{t}_2)=e^{\frac{\mathsf{t}_1}{6}}. 
\end{equation}
Using (\ref{pairing}), the $J$-function contains the following correlators:
\begin{equation}
  \label{deg18allowed}
  \langle \tau_1(e_{1,{\bf 1}})(e_{1,H})^{k_1}(e_{5,{\bf 1}})^{k_2}\rangle_{0,k_1+k_2+1}, \quad\langle(e_{1,H})^{k_1+1}(e_{5,{\bf 1}})^{k_2}\rangle_{0,k_1+k_2+1},\quad \langle(e_{1,H})^{k_1}(e_{5,{\bf 1}})^{k_2+1}\rangle_{0,k_1+k_2+1}.
\end{equation}
As in the previous example the base coordinates are not charged under the quantum symmetry, so that $q_B(\beta)=0$ and (\ref{rule2}) now reduces to
\begin{align}
  \label{deg18rule2}
  \text{(a)}:&\quad 4k_2=4\mod 6 \nonumber \\
  \text{(b)}:&\quad 4k_2=0\mod 6, 
\end{align}
where (a) holds for the first two types of correlators in (\ref{deg18allowed}) and (b) for the third type. In other words, a correlator of the form $\langle(e_{5,{\bf 1}})^{k_2}\rangle_{0,k_2,\beta}$ is only non-zero if $k_2=1,4,7,\ldots$. For $\beta=0$, the only three-point function compatible with the selection rules is $\langle (e_{1,H})^2e_{5,{\bf 1}}\rangle_{0,3,0}$. After making use of the dilaton and divisor equations, the $J$-function reduces to
\begin{align}
  J=&e_{1,{\bf 1}}+t_1e_{1,H}+t_2e_{5,{\bf 1}}\nonumber\\
  &+6e_{1,H^2}\left[\frac{t_1^2}{2}\langle (e_{1,H})^2e_{5,{\bf 1}}\rangle_{0,3,0}+\sum_{k_2,\beta\geq 0}e^{\beta t_1}\frac{t_2^{k_2}}{k_2!}\langle (e_{5,{\bf 1}})^{k_2+1}\rangle_{0,k_2+1,\beta} \right]\nonumber\\
  &+6e_{5,H}\left[t_1t_2\langle (e_{1,H})^2e_{5,{\bf 1}}\rangle_{0,3,0}+\sum_{k_2\geq0,\beta>0}\beta e^{\beta t_1}\frac{t_2^{k_2}}{k_2!}\langle (e_{5,{\bf 1}})^{k_2}\rangle_{0,k_2,\beta}\right]\nonumber\\
  &+6e_{5,H^2}\left[\frac{t_1^2}{2}t_2\langle (e_{1,H})^2e_{5,{\bf 1}}\rangle_{0,3,0}+\sum_{k_2,\beta\geq0}e^{\beta t_1}\frac{t_2^{k_2}}{k_2!}(-2+k_2+\beta t_1)\langle (e_{5,{\bf 1}})^{k_2}\rangle_{0,k_2,\beta}\right].
\end{align}
The coordinates $t_1,t_2$ are defined as
\begin{align}
  t_1=&\log u_1+\frac{575}{72}u_1+\frac{16764763u_1^2}{248832}+\frac{1962611009639 u_1^3}{2176782336}+\frac{196512288898979039 u_1^4}{13374150672384}\nonumber\\
  &+\frac{965367820745180163871 u_1^5}{3611020681543680}\nonumber\\
  &+\frac{\left(3165663842503837987968978413u_1^6+2105947261476274176 u_2^6\right)}{606512811305166962688}+\ldots\nonumber\\
  t_2=&u_2^3\left[\frac{1}{48}u_2+\frac{211u_1 u_2}{2592}+\frac{1242073u_1^2 u_2}{1119744}+\frac{255680653403u_1^3u_2}{13060694016}+\ldots\right]
\end{align}
We encounter a new phenomenon compared to the previous examples. Since the state defining the flat coordinate in the fibre direction is in a higher twisted sector, the leading term in the definition of $t_2$ is not linear in $u_2$. This poses some technical challenges and we have only been able to compute the expansion to low orders. In addition, the series expansions in $q_1=e^{t_1}$ and $q_2=t_2$ do not necessarily have integer exponents or rational coefficients. It is remarkable that, at least to the low orders we have computed, the $J$-function does not contain any fractional powers in the variables. For instance, we find
\begin{align}
  \frac{I_1^{H^2}}{I_1^{\bf 1}}=&\frac{1}{2} \log ^2 q_1+\frac{113}{18}  q_1+\frac{196319 
   q_1^2}{10368}+\frac{ \left(360744024241 q_1^3+1632586752q_2^3\right)}{3265173504}+\ldots
\end{align}
From this, we can read off
\begin{equation}
  \langle (e_{1,H})^2e_{5,{\bf 1}}\rangle_{0,3,0}=\frac{1}{6}
  \end{equation}
and for the non-zero higher invariants
\begingroup
\renewcommand{\arraystretch}{1.3}
\begin{equation}
  \begin{array}{c|ccccc}
    (n,\beta)&(1,1)&(1,2)&(1,3)&(4,0)&(4,1)\\
    \hline
    \langle(e_{5,{\bf 1}})^n\rangle_{0,n,\beta}&\frac{113}{108}&\frac{196319}{62208}&\frac{360744024241}{19591041024}&\frac{1}{2}&\frac{379}{72}
    \end{array}
\end{equation}
\endgroup

We end with a remark on the choice of coordinate $u_2$ for this model. The $\mathbb{Z}_6$-orbifold of the hybrid phase suggests to make the identification $u_2=e^{\frac{\mathsf{t_1}}{6}}$ with the GLSM parameter $\mathsf{t}_1$. However, the structure of the $I$-function also seems compatible with the choice $\tilde{u}_2=e^{\frac{\mathsf{t_1}}{3}}$. This affects the definition of the flat coordinates and will modify the values of any correlators $\langle(e_{5,{\bf 1}})^n\rangle_{0,n,\beta}$ with $n>4$. Using $\tilde{u}_2$ makes the calculation of the series expansions more tractable and we could confirm to higher orders that the $J$-function is polynomial and has rational coefficients with this choice of coordinates. Calling the modified expansion variables $\tilde{q}_1$ and $\tilde{q}_2$ we find
\begin{align}
  \frac{I_1^{H^2}}{I_1^{\bf 1}}=&\frac{1}{2} \log ^2 \tilde{q}_1+\frac{113}{18}  \tilde{q}_1+\frac{196319 
   \tilde{q}_1^2}{10368}+\frac{ \left(360744024241 \tilde{q}_1^3+1632586752
    \tilde{q}_2^3\right)}{3265173504}\nonumber\\
  &+\frac{ \left(2126024461302251 \tilde{q}_1^4+13200008085504
    \tilde{q}_2^3 \tilde{q}_1\right)}{2507653251072}\nonumber\\
  &+\frac{ \left(136566168805320679799
   \tilde{q}_1^5+1321621350440140800 \tilde{q}_2^3 \tilde{q}_1^2\right)}{18055103407718400}+\ldots
\end{align}
This gives the same invariants as in the table above. We can also extract higher invariants as follows:
\begingroup
\renewcommand{\arraystretch}{1.3}
\begin{equation}
  \begin{array}{c|ccc}
    (n,\beta)&(1,4)&(1,5)&(4,2)\\
    \hline
    \langle(e_{5,{\bf 1}})^n\rangle_{0,n,\beta}&\frac{2126024461302251}{15045919506432}&\frac{136566168805320679799}{108330620446310400}&\frac{9107167}{124416}
    \end{array}
\end{equation}
\endgroup
At this point we are not aware of a mechanism that matches the expansion variable of the $I$-function to the FI-theta parameters of the GLSM in a canonical way.  Therefore we cannot decide which expansion of the $J$-function is the correct one. To resolve this issue, one would have to derive the $I$-function directly from the hybrid theory. We leave this as an open problem.

\section{Outlook}
In this work we have analysed state spaces and extracted (conjectural) hybrid FJRW invariants for examples of good hybrid models with one and two K\"ahler parameters. We have shown that our results are consistent with mirror symmetry, results from supersymmetric localisation, and results from FJRW theory. Several open issues remain.

One important aspect is to further analyse the selection rule (\ref{rule2}) from both, the mathematics and physics perspective. The goal should be to give a precise formulation for all good hybrid models. In physics, this should come from a, potentially anomalous, quantum symmetry acting on the state space % remove if required by the referee 
or from a generalisation of the analysis of GLSM instanton moduli spaces given in \cite{Witten:1993yc}. From the mathematical perspective, one would require a more general analysis of the moduli spaces of multi-parameter hybrids. 

In this work we mostly focused on specific examples. Just as in the Landau-Ginzburg and the Calabi-Yau cases it would be desirable to have closed expressions for the $I$-functions of a larger class of hybrid models. To do so, having a classification of good hybrids similar to the classification of Landau-Ginzburg orbifolds \cite{Kreuzer:1992bi,Kreuzer:1992iq} seems useful. Most of the models that have been looked at in more detail in the literature so far have a $\mathbb{P}^n$-base and only one orbifold group acting that is compatible with $U(1)_V$. To get a comprehensive picture on hybrids, one should study models with more than one modulus in the base and several discrete groups acting. 

A related question is whether $J$-functions and FJRW invariants can also be defined in hybrid theories that are not good hybrids. While these pseudo-hybrids remain quite mysterious, some results \cite{Aspinwall:2009qy,Addington:2013gpa,Aspinwall:2017loy,Erkinger:2020cjt} imply that there is a lot of interesting structure and that $I$- and $J$-functions may be defined also for these models. These models seem to be the most generic phases of Calabi-Yau GLSMs. 

Another interesting direction is to gain a better understanding of possible connections between hybrid FJRW invariants and integer enumerative invariants. While one may not expect a connection to integer invariants in all cases, there are special examples of hybrids, like the model M1 discussed in Section~\ref{sec-m1} that behave ``almost'' like geometric models in large volume phases and can be interpreted in terms of branched double covers. Further examples of this type have also been found in \cite{Schimannek:2021pau} and in the context on non-abelian GLSMs \cite{Hori:2013gga}. It would be interesting to have a physics interpretation of the integer invariants. 

It might also be interesting to study the presence of any modular properties of the invariants of some hybrid examples. One candidate where one could expect traces of modularity is the two-parameter example of Section~\ref{sec-2parex2}. The large volume phase of this model is an elliptic fibration over $\mathbb{P}^2$. The hybrid phase can be understood as the elliptic fibre undergoing a transition to a Landau-Ginzburg orbifold phase while the base remains unaffected. One expects that this property is reflected in the invariants. This may also shed some light on the problem of choosing the correct $u$-coordinates that we found for this model.

One of the most challenging conceptual open questions is whether it is possible to relax the restriction to narrow sectors. Similar to the Landau-Ginzburg case, the narrow states as we have defined them for hybrids will not account for all states of the $(a,c)$-ring. For the definition of the correlators, it makes no difference whether a marginal state is narrow or, using the mathematical terminology, ``broad''. The situation is different when it comes to techniques to compute these invariants. There, the restriction to narrow states seems rather crucial. For example, the GLSM partition functions, as they are presently defined, seem to be blind to the broad states. In the Landau-Ginzburg case, it has been proposed in \cite{Knapp:2020oba}, paralleling considerations in geometry, to replace Landau-Ginzburg orbifold models with broad sectors by equivalent models where all the sectors are narrow. However, it is not clear to what extent these models are really equivalent. For instance, one should investigate whether the associated D-brane categories are equivalent. It would certainly be more satisfactory to incorporate the broad sectors into the existing framework. Interesting recent work in this direction can be found in \cite{Favero:2020cke}. 

While we have only focused on closed string invariants, one expects that the connections between different phases of GLSM can be formulated in terms of equivalences of D-brane categories. Similar to the Landau-Ginzburg/Calabi-Yau correspondence, there are more general correspondences involving hybrids that can be investigated using the GLSM in the spirit of \cite{Herbst:2008jq}. It would be interesting to work this out in concrete examples. See \cite{Zhao:2019dba,Chen:2020iyo,Guo:2021aqj} for interesting recent work in this direction. 

\bibliographystyle{utphys}
\bibliography{hybfjrw}
\end{document}